\documentclass[%
 reprint,
 amsmath,amssymb,
 aps,superscriptaddress,
]{revtex4-2}

\usepackage{graphicx}%
\usepackage{dcolumn}%
\usepackage{bm}%
\usepackage{siunitx} %
\usepackage{isomath}
\usepackage{soul}
\usepackage{color}
\usepackage[inkscapelatex=false]{svg}
\usepackage{xr-hyper}
\usepackage{hyperref}
\hypersetup{
    colorlinks = true,
    linkbordercolor = {green},
    allcolors = {blue},
}
\usepackage{xcolor,marvosym,moresize}

\usepackage{pifont}
\definecolor{magenta_tumbl}{RGB}{213, 101, 208}
\definecolor{blue_tumbl}{RGB}{64, 140, 202}
\definecolor{green_tumbl}{RGB}{31, 128, 31}
\definecolor{green_rots}{RGB}{32, 191, 129}
\definecolor{orange_spinn}{RGB}{255, 116, 70}
\definecolor{orange_rots}{RGB}{255, 100, 65}
\definecolor{violet_rots}{RGB}{109, 71, 159}

\definecolor{spinning}{RGB}{120, 81, 169}
\definecolor{tumbling}{RGB}{253, 112, 81}
\definecolor{tumbl_3}{RGB}{0, 199, 140}
\definecolor{tumbl_2}{RGB}{231, 184, 59}
\definecolor{dark_green_ratio}{RGB}{48, 79, 79}

\definecolor{green_e1x}{RGB}{96, 131, 32}
\definecolor{rosa_e2y}{RGB}{173, 99, 109}
\definecolor{gray_e2z}{RGB}{58, 78, 93}
\definecolor{squared_omega_y}{RGB}{253, 128, 129}
\definecolor{squared_omega_z}{RGB}{16, 105, 201}
\definecolor{rosa_omega_y}{RGB}{204, 74, 84}
\definecolor{blue_omega_z}{RGB}{61, 119, 171}

\begin{document}

\preprint{APS/123-QED}

\title{Full rotational dynamics of plastic microfibers in turbulence}

\author{Vlad Giurgiu}
\affiliation{Institute of Fluid Mechanics and Heat Transfer, TU Wien, 1060 Wien, Austria}

\author{Giuseppe Carlo Alp Caridi}
\affiliation{Institute of Fluid Mechanics and Heat Transfer, TU Wien, 1060 Wien, Austria}

\author{Marco De Paoli}%
\affiliation{Physics of Fluids Group, University of Twente, 7500AE Enschede, The Netherlands}
\affiliation{Institute of Fluid Mechanics and Heat Transfer, TU Wien, 1060 Wien, Austria}

\author{Alfredo Soldati}
\email{alfredo.soldati@tuwien.ac.at}
\affiliation{Institute of Fluid Mechanics and Heat Transfer, TU Wien, 1060 Wien, Austria}
\affiliation{Polytechnic Department, University of Udine, 33100 Udine, Italy}

\date{\today}

\begin{abstract}
The motion, settling, and dispersion of microplastics in the ocean are determined by their rotational dynamics. We present experiments on elongated, large aspect ratio, and mildly curved plastic fibers slightly longer than the Kolmogorov length scale. Exploiting their uniquely identifiable three dimensional orientation, we perform original optical Lagrangian investigations and provide a set of homogeneous data on their rotation rates around their longitudinal axis -- spinning rate -- and transversal axes -- tumbling rates -- which we explain in the context of the general features of turbulence.
\end{abstract}

\maketitle
Turbulent flows with suspended anisotropic particles are ubiquitous \cite{voth2017anisotropic,brandt2022particle} in many industrial \cite{lundell2011fluid,moffet2009situ} 
and natural processes \cite{sabban2011measurements,heymsfield1977precipitation,pedley1992hydrodynamic}.
The motion, dispersion, and sedimentation rate of anisotropic particles depend on their rotational dynamics. The recent and sudden increase of attention on the dynamics of such particles is driven by the growing concern about ocean microplastics pollution \cite{ross2021pervasive,ugwu2021microplastics,gruber2023waste,leslie2022discovery,chubarenko2016some,clark2023dispersion,tatsii2023shape}.
Most oceanic microplastics are small and elongated microfibers \cite{huntington2020first,kooi2019simplifying} and their dynamics depend on the forces and torques applied by the smallest turbulence scales \cite{parsa2012rotation,Brizzolara2021}. 
However, even if the statistical properties of turbulence at the smallest scale are rather well established and given that no closed form of the drag on non-spherical particles is available, a predictive understanding of the dynamics of anisotropic particles remains elusive. 
The full rotational dynamics of a small elongated particle is given by the rotation rates around the three principal axes: spinning rate, around the longitudinal axis and tumbling rates, around the two transversal axes.  
Current numerical \cite{byron2015shape,zhao2015rotation,marchioli2013rotation,marchioli2016relative,yang2018mean} and experimental investigations \cite{parsa2012rotation,baker2022experimental,shaik2023kinematics,hoseini2015finite} are limited to tumbling rates of small slender objects. 
The present study takes advantage of the specific shape of the investigated particles, which are microplastic slender fibers with aspect ratio $\sim 100$ and a slight curvature: these fibers behave like elongated ellipsoids \cite{voth2017anisotropic}, but their slight curvature provides the complete information on their three-dimensional orientation, 
and allows measurements of rotation rates around their three axes. 
We focus on wall-bounded turbulence, confirm previous  measurements of fiber tumbling obtained in different flow configurations, boundary layers and homogeneous isotropic turbulence (HIT), and discuss the results of each rotational component -- spinning and two tumbling rates.

\begin{figure}[b!]
\includegraphics[width=0.4\textwidth,trim={0cm 0cm 0cm 0cm},clip]{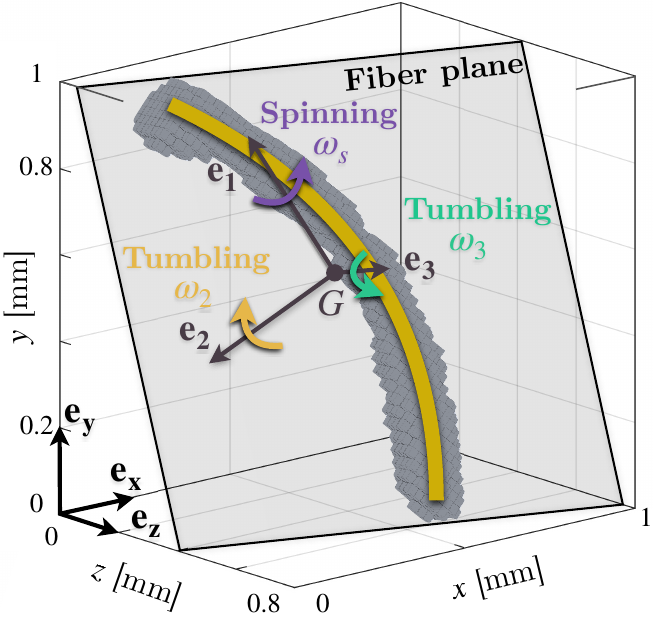}%
\caption{\label{fig:fibre}
A reconstructed fiber (dark gray voxels), its fitted polynomial (yellow line), the fiber-fixed axes ($\mathbf{e_1},\ \mathbf{e_2},\ \mathbf{e_3}$), and its center of mass $G$ (black dot) are shown.
Spinning ($\omega_s$) and tumbling rate components ($\omega_2$, $\omega_3$) around these axes are noted \cite{voth2017anisotropic}.
The laboratory reference frame is indicated by $\mathbf{e_x}$ (stream-wise), $\ \mathbf{e_y}$ (wall-normal), and $\mathbf{e_z}$ (span-wise).
}
\end{figure}

\begin{figure*}
\includegraphics[width=0.98\textwidth,trim={0cm 0cm 0cm 0cm},clip]{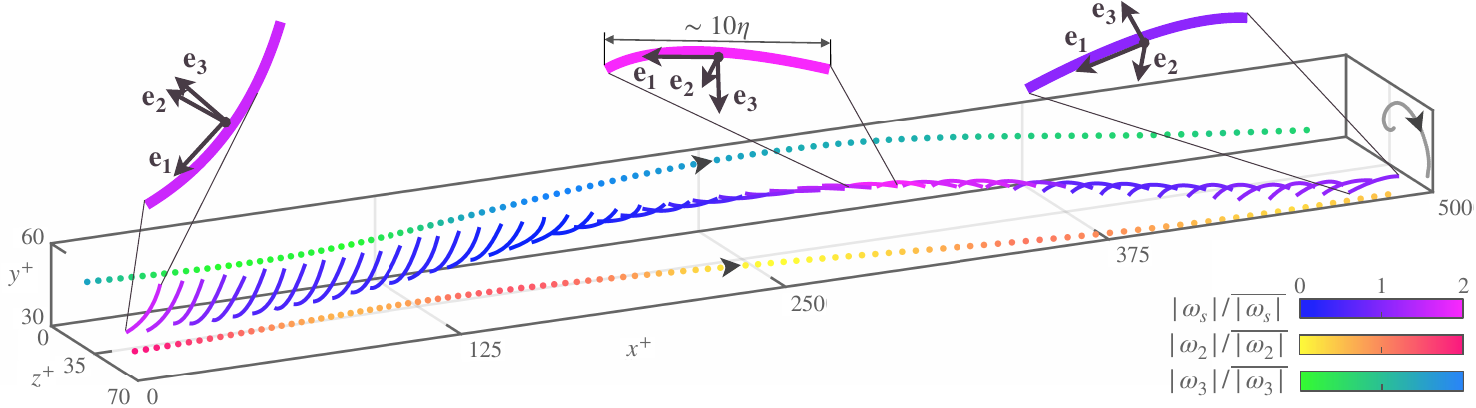}%
\caption{\label{fig:trajectory}
A 3D view of a fiber trajectory, showing its center of mass on $xz$, $xy$ (color indicates $|\omega_2|$, $|\omega_3|$), and $yz$ planes (gray line). 
The black arrows indicate its progression.
The fiber-equivalent polynomial is colored by $|\omega_s|$.
The rotation rates are normalized with their corresponding mean absolute track value of $|\omega_s|=\SI{1908}{\degree\per\second}$, $|\omega_2|=\SI{731}{\degree\per\second}$, and $|\omega_3|=\SI{665}{\degree\per\second}$. 
Above, three instances in time are magnified.
The black vectors ($\mathbf{e_1}$, $\mathbf{e_2},$ and $\mathbf{e_3}$) are the reference frame fixed to the fiber (see Fig.~\ref{fig:fibre}).
The fiber's end-to-end length relative to the Kolmogorov scale ($\eta$) near the wall is noted.
}
\end{figure*}

Experiments are performed in the TU Wien Turbulent Water Channel \cite{giurgiu2023tu},
at a shear Reynolds number $Re_{\tau} = u_{\tau}h/\nu=720$, where $u_{\tau}=\SI{20.5}{\milli\meter\per\second}$, $\nu = \SI{1.13}{\square\milli\meter\per\second}$, and $h=\SI{40}{\milli\meter}$ are the friction velocity, kinematic viscosity, and half-channel height, respectively.
Additional methodological details are provided in \cite{SupplMat_exp_det}.
In the fully developed turbulent channel flow, the viscous length and time scales are $\delta_\nu = \nu/u_{\tau} = \SI{55}{\micro\meter}$ and $\tau = \delta_{\nu}/u_{\tau} = \SI{2.7}{\milli\second}$, 
\cite{giurgiu2023tu}. 
The Kolmogorov length $[\eta=(\nu^3/\epsilon)^{1/4}]$ and time scales $[\tau_{\eta}=(\nu/\epsilon)^{1/2}]$ vary between wall and channel center in the range $\eta = 83-\SI{286}{\micro\meter}$ and $\tau_{\eta} = 5.7 - \SI{75}{\milli\second}$, respectively, where $\epsilon$ is the turbulent kinetic energy dissipation rate. 
To obtain the values for the Kolmogorov time and length scales we interpolated values of previous direct numerical simulations (DNS) at $Re_{\tau}=590$ \cite{moser1999direct} and $Re_{\tau}=950$ \cite{del2004scaling}. 
Fibers have a density of $\rho_{f}=\SI{1.15}{\gram\per\cubic\centi\meter}$ (see \cite{SupplMat_settling} and \cite{bhowmick2024inertia} for details about the settling behavior), are $l=\SI{1.2}{\milli\meter}$ long and $d=\SI{10}{\micro\meter}$ in diameter (aspect ratio $\lambda=l/d=120$) \cite{flock}. 
Relative to the local Kolmogorov length scale, $l$ ranges between $4.2\eta$ and $11.4\eta$, from the channel center to $\SI{2}{\milli\meter}$ above the wall, respectively. 
At $Re_{\tau}=720$, they are rigid and inertia-less with a translational Stokes number \cite{bernstein1994direct} $St\sim \mathcal{O}(10^{-2})$ \cite{alipour2022influence}.
The tumbling Stokes number \cite{bounoua2018tumbling} is $St_t \sim \mathcal{O}(10^{-5})$.
Fibers are dispersed in water to form a dilute suspension with a volume fraction of $\mathcal{O}(10^{-9})$, with negligible inter-particle interactions and influence on the flow \cite{brandt2022particle}. 
Measurements are performed at two wall-normal locations, $y^+=y/\delta_{\nu} < 270$ and $640<y^+<800$.
The laser-illuminated fibers are imaged with six high-speed cameras \cite{giurgiu2023tu} 
to reconstruct their shape and measure their dynamics \cite{alipour2021long}. In Fig.~\ref{fig:fibre}, we show as dark gray voxels the tomographic reconstruction of the light scattered by one fiber (see ~\cite{SupplMat_method}).
Fibers are contained in a plane (light gray plane)\cite{alipour2021long,alipour2022influence}, with 
their center of mass ($G$) and eigenvectors of their inertia tensor ($\mathbf{e_1}$, $\mathbf{e_2}$, $\mathbf{e_3}$ -- in order of ascending eigenvalues) defining their position and orientation. 
A $2$nd order polynomial (yellow line in Fig.~\ref{fig:fibre}) is fitted to the reconstructed fiber to measure its curvature.
All statistics have been computed with fibers having a curvature in a limited range.
Consult \cite{SupplMat_Fig_curv_pdf} for the distribution of the fibers' curvature and the considered range.
In the fiber reference frame, the spinning rate, $\omega_s$ is defined as the rotation rate around $\mathbf{e_1}$ and the two tumbling rate components, $\omega_2$ and $\omega_3$ as the rotation rates around $\mathbf{e_2}$ and $\mathbf{e_3}$, respectively
\cite{voth2017anisotropic}. In the laboratory reference frame, rotation rates around $\mathbf{e_x}$ (stream-wise), $\mathbf{e_y}$ (wall-normal), and $\mathbf{e_z}$ (span-wise) directions are $\omega_x$, $\omega_y$, and $\omega_z$, respectively (see \cite{SupplMat_rotation_rates_comp}). 

The richness of the present measurements is demonstrated by a fiber trajectory near the wall in Fig.~\ref{fig:trajectory}. 
This fiber moves $\mathcal{O}(500 \delta_{\nu})$ in the stream-wise ($x^+$) and $\mathcal{O}(30\delta_{\nu})$ in the wall-normal ($y^+$) and span-wise ($z^+$) directions. Its trajectory is colored by its spinning rate scaled by the track average (denoted with $\overline{\ \cdot \ }$). 
The fiber is shown magnified at three different times to illustrate the temporal change of its orientation, defined by the three vectors $\mathbf{e_1}$, $\mathbf{e_2}$, and $\mathbf{e_3}$. 
The traces of the fiber's center of mass onto the Cartesian planes are also shown, with those onto planes $xz$ and $xy$ being colored by $\omega_2$ and $\omega_3$, respectively and scaled by their track average.

\begin{figure}
\includegraphics[width=0.48\textwidth,trim={0cm 0cm 0cm 0cm},clip]{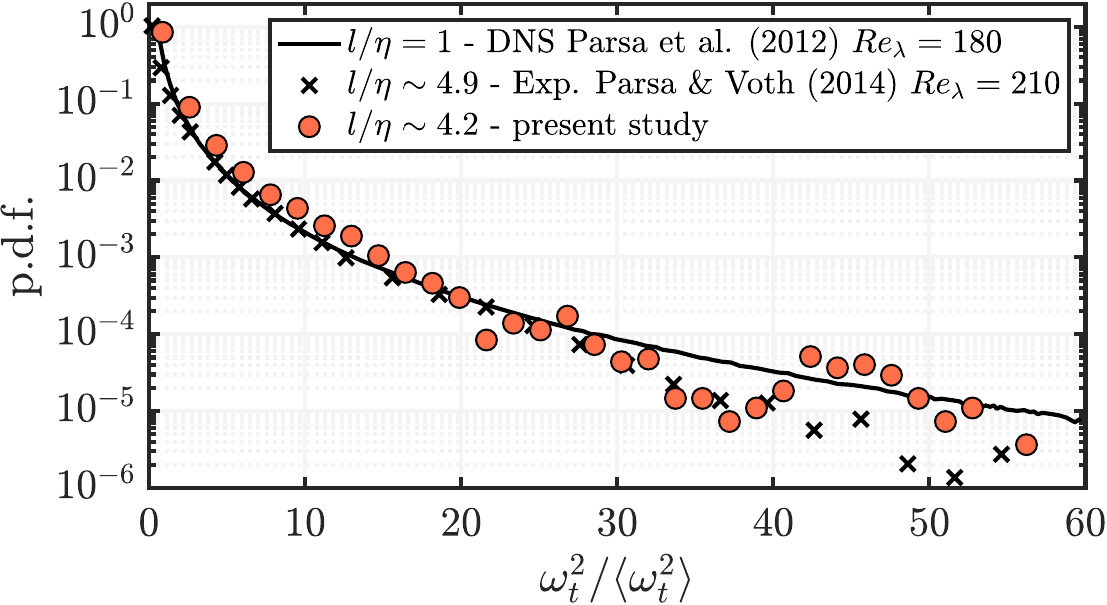}
\caption{\label{fig:tumbl_pdf} 
P.d.f. of the normalized tumbling rate measured in the center of the channel (\textcolor{tumbling}{$\bullet$}) compared to numerical simulations of straight rods with $l/\eta=1$ (\cite{parsa2012rotation}, black line) and experiments $l/\eta\sim 4.9$ (\cite{parsa2014inertial}, \bm{$\times$}) in HIT. $Re_{\lambda}$ is the Taylor Reynolds number.}
\end{figure}

The length and time scales of turbulence increase from the wall towards the channel center, where turbulence features are similar to HIT. 
We focus first on the channel center region.
In Fig.~\ref{fig:tumbl_pdf} we show the p.d.f. of the squared tumbling rate $\omega_t^2 = \omega_2^2 + \omega_3^2$ normalized by its average value, $\langle\omega_t^2\rangle$, computed over the entire dataset and indicated here by angular brackets. 
The comparison with measurements ($l/\eta=4.9$) \cite{parsa2014inertial} and direct numerical simulations (DNS) of rods ($l/\eta=1$) \cite{parsa2012rotation} shows excellent agreement, particularly in the distribution's long tails, where rare, strong rotation rate events occur. 
DNS data exhibit slightly larger rotation rates at high values of normalized squared tumbling rates. 
This is possibly due to the effect of the fiber's finite length in the experiments and the assumption of tracing, point-wise fibers in DNS.
Consult \cite{SupplMat_Fig_tumbl_pdf_number_sample} for the effect of sample size on the p.d.f. of tumbling rates.
The slight curvature of fibers appears to have no discernible effect on the p.d.f. of tumbling rates. 
Consult \cite{SupplMat_Fig_curv_effect} for the effect of curvature on rotation rates.

\begin{figure}
\includegraphics[width=0.48\textwidth,trim={0cm 0cm 0cm 0cm},clip]{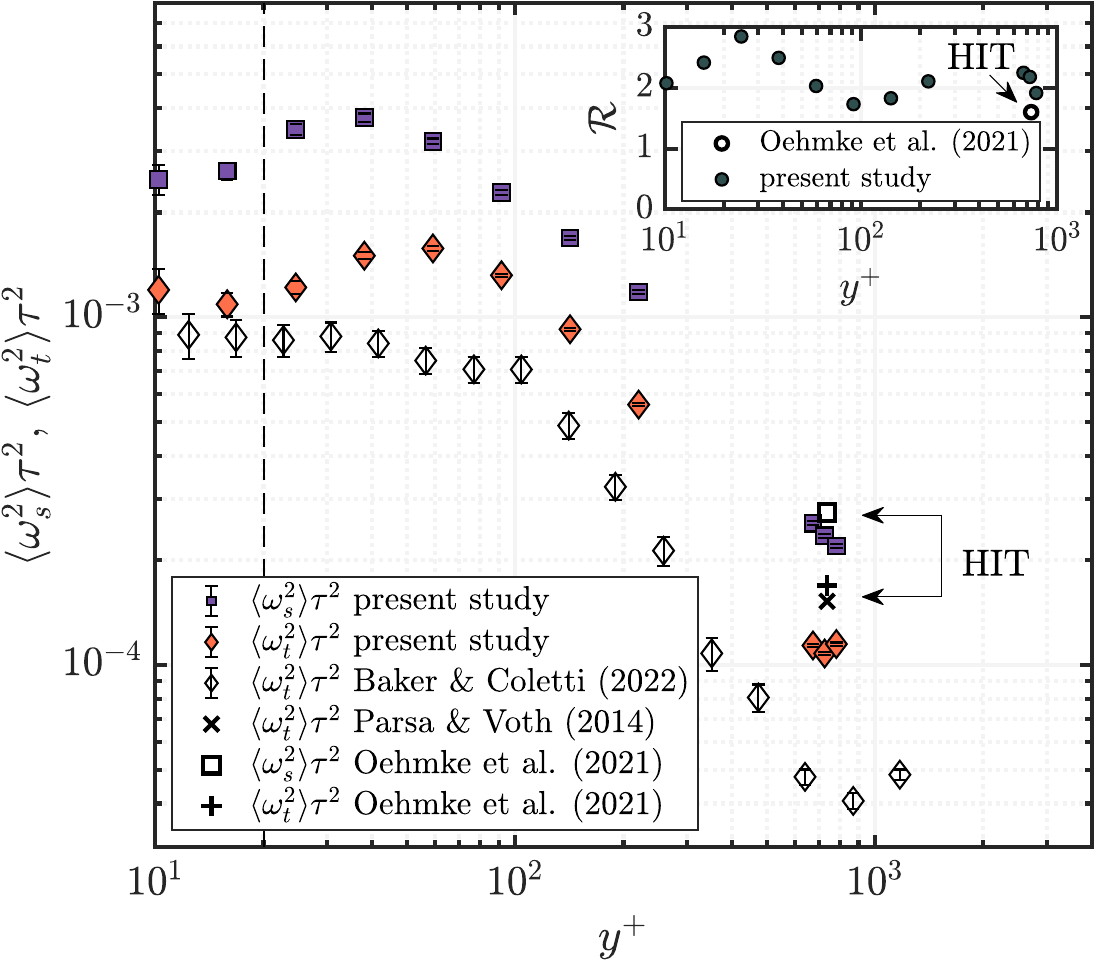}
\caption{\label{fig:tumbl_spinn} 
Main panel: mean square tumbling and spinning rates normalized by the viscous time-scale over the wall-normal coordinate. 
Combined tumbling in the present study ($l/\eta\approx 4.2$, \textcolor{tumbling}{$\blacklozenge$}) and straight rods in a turbulent boundary layer (\cite{baker2022experimental}, $l/\eta\approx 22-42$, $Re_{\tau}=620$, $St=6$, $\diamondsuit$).
At the channel center, tumbling measurements of straight rods in HIT $l/\eta\approx 4.9$ (\cite{parsa2014inertial}, $Re_{\lambda}=210$, \bm{$\times$}) and $l/\eta\approx 11$ (\cite{oehmke2021spinning}, $Re_{\lambda}=630$, $\mathbf{+}$).
The fiber length ($L_f^+\approx20$, $- -$) is displayed. 
Spinning rate for present experiments ($l/\eta\approx 4.2$, \textcolor{spinning}{$\blacksquare$}) and straight rods in HIT (\cite{oehmke2021spinning}, $l/\eta\approx 11$, $Re_{\lambda}=630$, \Squarepipe).
Error-bars represent confidence intervals at $95\%$ and $99\%$ confidence level for data from Ref.~\cite{baker2022experimental} and present study, respectively. 
Note: the values for HIT (\bm{$\times$}, $\mathbf{+}$, and \Squarepipe) are obtained by multiplying the reported values for $\langle \omega_t^2\rangle\tau_{\eta}^2$ of $0.12$ \cite{parsa2014inertial} and $0.13$ \cite{oehmke2021spinning} and for $\langle \omega_s^2\rangle\tau_{\eta}^2$ of $0.21$ \cite{oehmke2021spinning} with the ratio $\tau^2 / \tau_{\eta}^2=1.6\times10^{-7}$ (here $\tau_{\eta}$ is the mean Kolmogorov time scale in the channel central region).
Inset: ratio of the mean squared spinning to mean squared tumbling rate $\mathcal{R} = \langle \omega_s^2\rangle/\langle \omega_t^2\rangle$ for the present study ($l/\eta\approx 4.2$, \textcolor{dark_green_ratio}{$\bullet$}) over the wall-normal coordinate. 
At the channel center, the measured ratio for HIT (\cite{oehmke2021spinning}, $l/\eta\approx 11$, $Re_{\lambda}=630$, \Circpipe). Arrows indicate measurements in HIT configuration.
}
\end{figure}
Previous observations have shown that the tumbling rate is a strong function of the wall distance, with fibers tumbling faster near the wall \cite{baker2022experimental,alipour2022influence,shaik2023kinematics,shaik2020measurements,marchioli2013rotation,mortensen2008dynamics}. 
In the main panel of Fig.~\ref{fig:tumbl_spinn}, the mean squared tumbling rate normalized by the viscous time-scale as a function of the $\delta_{\nu}$--scaled wall distance is shown (see \cite{SupplMat_Fig_convergence}).
The present data are compared with measurements in a boundary layer at $Re_\tau = 620$, with fibers three times longer than the present ones  \cite{baker2022experimental}.  
As a reference, we show measurements in HIT of fibers one order of magnitude longer than $\eta$ \cite{oehmke2021spinning,parsa2014inertial} and with maximum aspect ratio of $\approx 10$. 
Present values of tumbling rates exhibit the same qualitative behavior of previous data \cite{baker2022experimental}, with values higher in the wall region
and decreasing towards the center of the channel.
The high tumbling rates compared to Ref.~\cite{baker2022experimental} may be due to the shorter length of our fibers ($l/\eta \lesssim 11$ - approximately in the dissipative range of turbulence \cite{parsa2014inertial}) compared to rods in Ref.~\cite{baker2022experimental} ($22 \lesssim l/\eta \lesssim 42 $ - in the inertial sub-range, where tumbling rates scale with $\left(l/\eta\right)^{-4/3}$ \citep{parsa2014inertial,oehmke2021spinning}).
The different tumbling rates compared to Ref.~\cite{baker2022experimental} may not be explained by differences in the translational Stokes number.
At $y^+=50$, Ref.~\cite{marchioli2013rotation} found negligible differences in the root-mean-square of rotation rates between tracer particles ($St=0$) and rigid fibers ($\lambda=50$, $l/\eta\approx 10$) at two different Stokes numbers ($St=1$ and $5$).
Furthermore, Ref.~\cite{zhao2015rotation} reported a reduction of only $\approx 20\%$ in the mean squared tumbling rate between tracer particles ($St=0$) and inertial straight fibers ($St=30$, $\lambda=50$) at the channel center.
The mean squared tumbling rate exhibits a peak at $y^+\approx 60$, which has not been observed in previous experiments. 
At the channel center, values similar to previous HIT experimental data are recovered \cite{parsa2014inertial,oehmke2021spinning}. 
Previous experiments on fibers in wall-bounded and HIT turbulence are limited to measuring tumbling rates only. 
Notable exceptions are Ref.~\cite{oehmke2021spinning} (tumbling and spinning rates of cylinders) and Ref.~\cite{marcus2014measurements} (solid body rotation rates of jacks and crosses).
Examples also include numerical studies with prolate and oblate ellipsoids modeled as point-particles and carried out by Ref.~\cite{byron2015shape} (all rotation rates) and Ref.~\cite{zhao2015rotation} (tumbling and spinning rates).
Our measured spinning rates at the channel center are found in good agreement (within $20\%$) with those of longer, lower aspect ratio, straight rods in HIT \cite{oehmke2021spinning,parsa2014inertial}.
While the spinning rate trends qualitatively similar to tumbling, its peak occurs at $y^+ \approx 40$.
The inset of Fig.~\ref{fig:tumbl_spinn} displays the ratio ($\mathcal{R}$) of the mean squared spinning to mean squared tumbling rates as a function of the wall-normal coordinate and is compared with measurements in HIT \cite{oehmke2021spinning}.
$\mathcal{R}$ consistently exceeds unity, reaching up to three (at $y^+\approx 25$).
Near the channel center, the ratio agrees fairly with HIT experiments \cite{oehmke2021spinning}. 
This corroborates other studies \cite{zhao2015rotation,byron2015shape,pujara2021shape} that also reported spinning rates higher than tumbling rates.
This was attributed by Ref.~\cite{oehmke2021spinning} to rods being preferentially trapped in elongated vortical structures, which have been described by Refs.~\cite{douady1991direct,jimenez1998characteristics} in HIT.
In wall bounded flows, such coherent structures, are typically oriented in the stream-wise direction \cite{robinson1991coherent,adrian2007hairpin,yang2011geometric,green2007detection,jimenez2018coherent,sharma2013coherent}.
In Ref.~\cite{zhao2015rotation}, in agreement with Refs.~\cite{challabotla2015shape,zhao2016spheroids,yang2018mean,baker2022experimental,shaik2020measurements,shaik2023kinematics,alipour2022influence}, a high probability of a stream-wise alignment of the longitudinal axis of inertia-less rods ($St=0$) near the wall (at $y^+\approx 10$) was observed.
Considering the perpendicular alignment of rods to vorticity (mainly oriented in $\mathbf{e_z}$), it was argued by Ref.~\cite{zhao2015rotation} that the mean shear cannot explain tumbling rates weaker than spinning rates, and this was attributed to the transport and interaction of inertia-less rods with coherent structures, resulting in the preferential stream-wise alignment of rods. 

\begin{figure}
\includegraphics[width=0.48\textwidth,trim={0cm 0cm 0cm 0cm},clip]{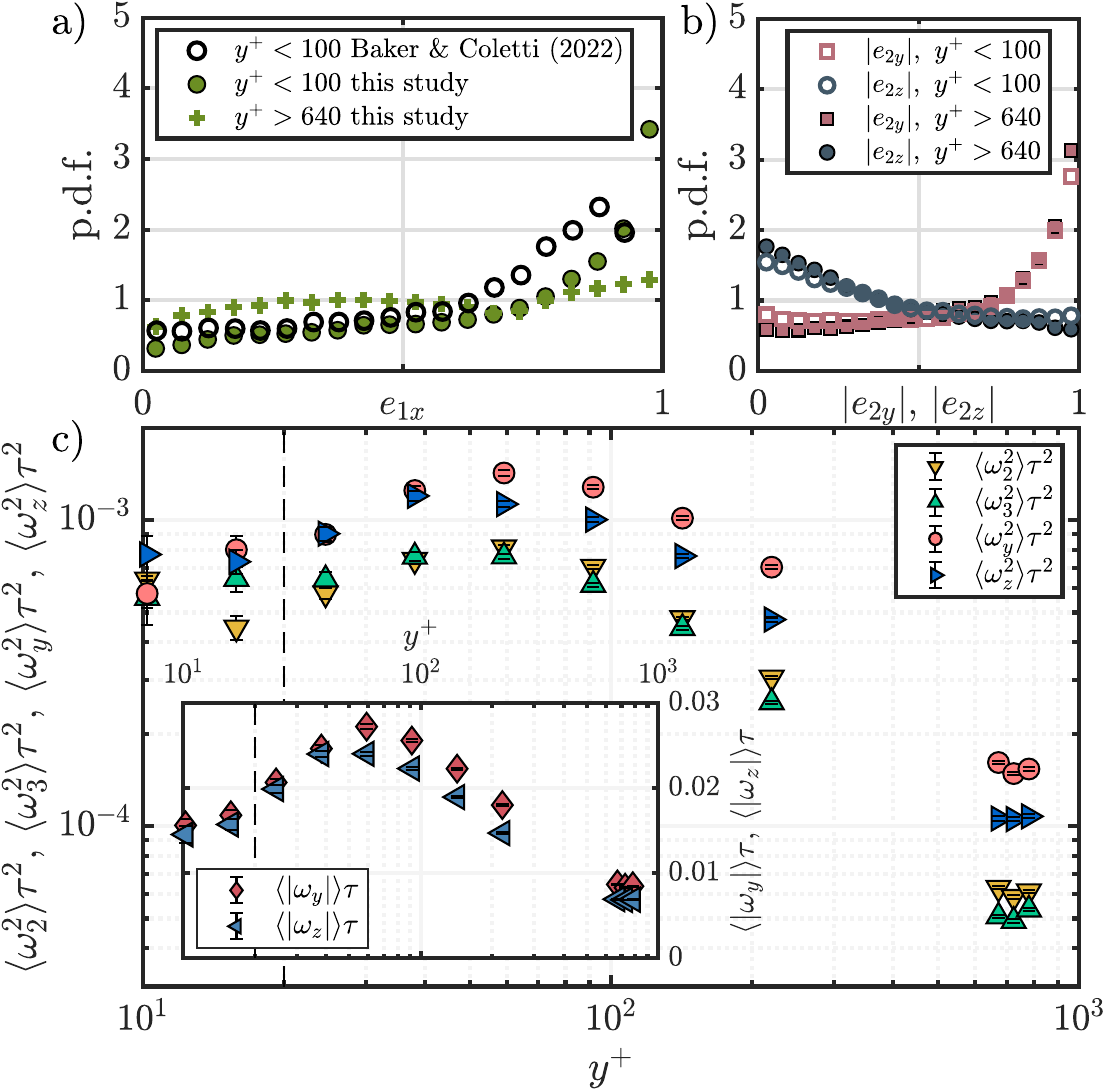}
\caption{\label{fig:orient_tumbl_vort}
Panel a): p.d.f. of the stream-wise component of the longitudinal axis vector ($e_{1x}$) in the present study for $y^+<100$ ($\mathbf{\textcolor{green_e1x}{\bullet}}$) and $640<y^+<800$ (\textcolor{green_e1x}{+}) compared to straight rods in a turbulent boundary layer (\cite{baker2022experimental}, \Circpipe). 
Panel b): p.d.f. of the absolute wall-normal ($e_{2y}$ for $y^+<100$, \textcolor{rosa_e2y}{\Squarepipe} and $640<y^+<800$, \small\textcolor{rosa_e2y}{$\blacksquare$}) and span-wise ($e_{2z}$ for $y^+<100$, \textcolor{gray_e2z}{\Circpipe} and $640<y^+<800$, \textcolor{gray_e2z}{$\bullet$}) components of the transversal axis vector $\mathbf{e_2}$. 
Main panel of c): mean squared components of tumbling scaled by the viscous time: second (\textcolor{tumbl_2}{$\blacktriangledown$}) and third (\textcolor{tumbl_3}{$\blacktriangle$}) over $y^+$. Mean squared wall-normal (\small\textcolor{squared_omega_y}{$\bullet$}) and span-wise (\textcolor{squared_omega_z}{$\blacktriangleright$}) rotation rates over $y^+$. 
Inset of panel c): mean absolute wall-normal (\textcolor{rosa_omega_y}{$\blacklozenge$}) and span-wise (\small\textcolor{blue_omega_z}{$\blacktriangleleft$}) rotation rates over $y^+$. 
Error-bars represent confidence intervals at $99\%$ confidence level.
}
\end{figure}
Advancing previous studies, in which only the orientation of the longitudinal axis was measured, this study investigates the orientation of all fiber axes: $\mathbf{e_1}$, $\mathbf{e_2}$, and $\mathbf{e_3}$.
In Fig.~\ref{fig:orient_tumbl_vort}a we show the p.d.f. of the stream-wise component ($e_{1x}$) of $\mathbf{e_1}$, of our fibers and that of straight rods ~\cite{baker2022experimental}. 
We consider two flow regions.
In the near-wall region (here $y^+ < 100$), the p.d.f. peaks near unity, indicating a preferential alignment of the longitudinal fiber axis with the stream-wise direction ($\mathbf{e_1}\approx \mathbf{e_x}$).
While a fair agreement with Ref.~\cite{baker2022experimental} is observed, our measured stream-wise alignment is stronger, as evidenced by the narrower distribution.
The stream-wise orientation near the wall ($y^+<50$) has been also observed in numerical studies \cite{do2014simulation,cui2021alignment}.
This preferential orientation is attributed to rods aligning with the strongest Lagrangian stretching direction and was observed in wall-bounded \cite{zhao2016spheroids,yang2018mean} and homogeneous isotropic turbulence \cite{ni2014alignment}.
At the channel center (here $640<y^+<800$) and consistent with previous studies \cite{zhao2015rotation,baker2022experimental,shaik2020measurements}, the probability of alignment is less than half of that near the wall.
In Fig.~\ref{fig:orient_tumbl_vort}b we show the p.d.f. of the absolute wall-normal ($e_{2y}$) and span-wise ($e_{2z}$) components of $\mathbf{e_2}$ in the same regions as in Fig.~\ref{fig:orient_tumbl_vort}a. 
In both regions, a preferential alignment with the wall-normal direction ($\mathbf{e_2}\approx \mathbf{e_y}$) is observed, as indicated by the two peaks of $|e_{2y}|$ and the low probability of $|e_{2z}|$ close to unity. 
This indicates that in the near wall-region the plane of the curved fibers considered is most probably parallel to the $x-y$ plane.
In \cite{SupplMat_Fig_curv_effect_on_orientation} we quantify the effect of fibre curvature on fiber orientation and we show that in the curvature range examined this effect is negligible.
The near-wall preferential orientation of the fibers is used to investigate the effect on their rotation of the near wall mean shear -- and in turn, of the mean vorticity, mainly oriented in the span-wise direction \cite{iwamoto2002reynolds}. 
To this aim, we use the fiber rotation rates measured in the fiber-fixed frame and in the laboratory frame of reference.
In the main panel of Fig.~\ref{fig:orient_tumbl_vort}c we show, as a function of $y^+$, the mean squared rotation rates around the two fiber transversal axes, $\langle \omega_2^2 \rangle \tau^2$ and $\langle \omega_3^2 \rangle \tau^2$ and around the wall-normal and span-wise directions, $\langle \omega_y^2 \rangle \tau^2$ and $\langle \omega_z^2 \rangle \tau^2$, respectively. 
Across the channel height, all rates follow nearly the same trend with a minimum in the central region and a peak at $y^+ = 60$.
We observe that values for $\langle \omega_y^2 \rangle \tau^2$ and $\langle \omega_z^2 \rangle \tau^2$ are similar, and so are the the values of $\langle \omega_2^2 \rangle\tau^2$ and $ \langle \omega_3^2 \rangle \tau^2$. In connection with the results on fiber orientation discussed before, this demonstrates no prevalent role of the mean shear/mean vorticity near the wall. 
As further evidence, in the inset of Fig.~\ref{fig:orient_tumbl_vort}c we show the mean absolute rotation rates around the wall-normal ($\omega_y$) and span-wise directions ($\omega_z$) as a function of $y^+$: both exhibit similar magnitudes across the channel height, further demonstrating no major influence of the span-wise oriented mean shear.
In a DNS study \cite{zhao2019mapping}, it was observed that the rotational mechanical energy of inertia-less ($St=0$), high aspect ratio ($\lambda=10$) ellipsoids was dominated by angular velocity fluctuations rather than by the mean angular velocity.
The former were associated with turbulent fluctuations, while the latter with the mean shear.
Conversely, the rotational mechanical energy of inertial ellipsoids ($St=30$, $\lambda = 0.1 - 10$) was found to be  dominated by the mean angular velocity, so by the mean shear. 
This is also supported by experimental evidence \cite{baker2022experimental},  where the high tumbling rates of inertial rods ($St=6$, $\lambda=11.8$) were attributed to the mean shear.

With the aim of characterizing the dispersion and sedimentation rates of microplastics in oceanic turbulence, we analyzed the rotation rates of slightly curved, inertia-less fibers in turbulent channel flow. 
Their shape enables measurements of their full rotational dynamics both in the fiber and laboratory frame of reference. 
Our findings align with those in both homogeneous isotropic \cite{parsa2014inertial,oehmke2021spinning} and wall-bounded turbulence studies \cite{baker2022experimental}. 
We find that the rotation rates of inertia-less fibers are dominated by turbulent fluctuations and not by the mean shear,  
that spinning rates are higher than tumbling rates \cite{oehmke2021spinning,zhao2015rotation,zhao2019mapping}, and that the components of tumbling rates have similar values, and also the spanwise and wall-normal components of the rotation rates have similar values.

Our technique may be used in the future to study the interaction between fibers and a solid boundary. Future studies might employ simultaneous measurements of curved fibers and surrounding flow to examine how near-wall structures influence the alignment of the fiber plane with the $x-y$ plane.

\begin{acknowledgments}
The authors acknowledge the Mr. F. Neuwirth and Mr. W. Jandl for the support in design and construction of the experimental facility. 
V.G. acknowledges the financial support provided by FSE S3 HEaD (Grant No. 1619942002). 
This research was funded in part by the Austrian Science Fund (FWF) (Grant No. P-35505). V.G. and A.S. also gratefully acknowledge funding from the PRIN project “Advanced computations and experiments in turbulent multiphase flow” (Project No. 2017RSH3JY)
\end{acknowledgments}

\bibliography{bibliography}

\providecommand{\noopsort}[1]{}\providecommand{\singleletter}[1]{#1}%
\begin{thebibliography}{71}%
\makeatletter
\providecommand \@ifxundefined [1]{%
 \@ifx{#1\undefined}
}%
\providecommand \@ifnum [1]{%
 \ifnum #1\expandafter \@firstoftwo
 \else \expandafter \@secondoftwo
 \fi
}%
\providecommand \@ifx [1]{%
 \ifx #1\expandafter \@firstoftwo
 \else \expandafter \@secondoftwo
 \fi
}%
\providecommand \natexlab [1]{#1}%
\providecommand \enquote  [1]{``#1''}%
\providecommand \bibnamefont  [1]{#1}%
\providecommand \bibfnamefont [1]{#1}%
\providecommand \citenamefont [1]{#1}%
\providecommand \href@noop [0]{\@secondoftwo}%
\providecommand \href [0]{\begingroup \@sanitize@url \@href}%
\providecommand \@href[1]{\@@startlink{#1}\@@href}%
\providecommand \@@href[1]{\endgroup#1\@@endlink}%
\providecommand \@sanitize@url [0]{\catcode `\\12\catcode `\$12\catcode
  `\&12\catcode `\#12\catcode `\^12\catcode `\_12\catcode `\%12\relax}%
\providecommand \@@startlink[1]{}%
\providecommand \@@endlink[0]{}%
\providecommand \url  [0]{\begingroup\@sanitize@url \@url }%
\providecommand \@url [1]{\endgroup\@href {#1}{\urlprefix }}%
\providecommand \urlprefix  [0]{URL }%
\providecommand \Eprint [0]{\href }%
\providecommand \doibase [0]{https://doi.org/}%
\providecommand \selectlanguage [0]{\@gobble}%
\providecommand \bibinfo  [0]{\@secondoftwo}%
\providecommand \bibfield  [0]{\@secondoftwo}%
\providecommand \translation [1]{[#1]}%
\providecommand \BibitemOpen [0]{}%
\providecommand \bibitemStop [0]{}%
\providecommand \bibitemNoStop [0]{.\EOS\space}%
\providecommand \EOS [0]{\spacefactor3000\relax}%
\providecommand \BibitemShut  [1]{\csname bibitem#1\endcsname}%
\let\auto@bib@innerbib\@empty
\bibitem [{\citenamefont {Voth}\ and\ \citenamefont
  {Soldati}(2017)}]{voth2017anisotropic}%
  \BibitemOpen
  \bibfield  {author} {\bibinfo {author} {\bibfnamefont {G.~A.}\ \bibnamefont
  {Voth}}\ and\ \bibinfo {author} {\bibfnamefont {A.}~\bibnamefont {Soldati}},\
  }\bibfield  {title} {\bibinfo {title} {Anisotropic particles in turbulence},\
  }\href@noop {} {\bibfield  {journal} {\bibinfo  {journal} {Annu. Rev. Fluid
  Mech.}\ }\textbf {\bibinfo {volume} {49}},\ \bibinfo {pages} {249} (\bibinfo
  {year} {2017})}\BibitemShut {NoStop}%
\bibitem [{\citenamefont {Brandt}\ and\ \citenamefont
  {Coletti}(2022)}]{brandt2022particle}%
  \BibitemOpen
  \bibfield  {author} {\bibinfo {author} {\bibfnamefont {L.}~\bibnamefont
  {Brandt}}\ and\ \bibinfo {author} {\bibfnamefont {F.}~\bibnamefont
  {Coletti}},\ }\bibfield  {title} {\bibinfo {title} {Particle-laden
  turbulence: progress and perspectives},\ }\href@noop {} {\bibfield  {journal}
  {\bibinfo  {journal} {Annu. Rev. Fluid Mech.}\ }\textbf {\bibinfo {volume}
  {54}},\ \bibinfo {pages} {159} (\bibinfo {year} {2022})}\BibitemShut
  {NoStop}%
\bibitem [{\citenamefont {Lundell}\ \emph {et~al.}(2011)\citenamefont
  {Lundell}, \citenamefont {S{\"o}derberg},\ and\ \citenamefont
  {Alfredsson}}]{lundell2011fluid}%
  \BibitemOpen
  \bibfield  {author} {\bibinfo {author} {\bibfnamefont {F.}~\bibnamefont
  {Lundell}}, \bibinfo {author} {\bibfnamefont {L.~D.}\ \bibnamefont
  {S{\"o}derberg}},\ and\ \bibinfo {author} {\bibfnamefont {P.~H.}\
  \bibnamefont {Alfredsson}},\ }\bibfield  {title} {\bibinfo {title} {Fluid
  mechanics of papermaking},\ }\href@noop {} {\bibfield  {journal} {\bibinfo
  {journal} {Annu. Rev. Fluid Mech.}\ }\textbf {\bibinfo {volume} {43}},\
  \bibinfo {pages} {195} (\bibinfo {year} {2011})}\BibitemShut {NoStop}%
\bibitem [{\citenamefont {Moffet}\ and\ \citenamefont
  {Prather}(2009)}]{moffet2009situ}%
  \BibitemOpen
  \bibfield  {author} {\bibinfo {author} {\bibfnamefont {R.~C.}\ \bibnamefont
  {Moffet}}\ and\ \bibinfo {author} {\bibfnamefont {K.~A.}\ \bibnamefont
  {Prather}},\ }\bibfield  {title} {\bibinfo {title} {In-situ measurements of
  the mixing state and optical properties of soot with implications for
  radiative forcing estimates},\ }\href@noop {} {\bibfield  {journal} {\bibinfo
   {journal} {Proc Natl Acad Sci U S A}\ }\textbf {\bibinfo {volume} {106}},\
  \bibinfo {pages} {11872} (\bibinfo {year} {2009})}\BibitemShut {NoStop}%
\bibitem [{\citenamefont {Sabban}\ and\ \citenamefont {van
  Hout}(2011)}]{sabban2011measurements}%
  \BibitemOpen
  \bibfield  {author} {\bibinfo {author} {\bibfnamefont {L.}~\bibnamefont
  {Sabban}}\ and\ \bibinfo {author} {\bibfnamefont {R.}~\bibnamefont {van
  Hout}},\ }\bibfield  {title} {\bibinfo {title} {Measurements of pollen grain
  dispersal in still air and stationary, near homogeneous, isotropic
  turbulence},\ }\href@noop {} {\bibfield  {journal} {\bibinfo  {journal} {J.
  Aerosol Sci.}\ }\textbf {\bibinfo {volume} {42}},\ \bibinfo {pages} {867}
  (\bibinfo {year} {2011})}\BibitemShut {NoStop}%
\bibitem [{\citenamefont {Heymsfield}(1977)}]{heymsfield1977precipitation}%
  \BibitemOpen
  \bibfield  {author} {\bibinfo {author} {\bibfnamefont {A.~J.}\ \bibnamefont
  {Heymsfield}},\ }\bibfield  {title} {\bibinfo {title} {{Precipitation
  development in stratiform ice clouds: {A} microphysical and dynamical
  study}},\ }\href@noop {} {\bibfield  {journal} {\bibinfo  {journal} {J.
  Atmos. Sci.}\ }\textbf {\bibinfo {volume} {34}},\ \bibinfo {pages} {367}
  (\bibinfo {year} {1977})}\BibitemShut {NoStop}%
\bibitem [{\citenamefont {Pedley}\ and\ \citenamefont
  {Kessler}(1992)}]{pedley1992hydrodynamic}%
  \BibitemOpen
  \bibfield  {author} {\bibinfo {author} {\bibfnamefont {T.}~\bibnamefont
  {Pedley}}\ and\ \bibinfo {author} {\bibfnamefont {J.~O.}\ \bibnamefont
  {Kessler}},\ }\bibfield  {title} {\bibinfo {title} {Hydrodynamic phenomena in
  suspensions of swimming microorganisms},\ }\href@noop {} {\bibfield
  {journal} {\bibinfo  {journal} {Annu. Rev. Fluid Mech.}\ }\textbf {\bibinfo
  {volume} {24}},\ \bibinfo {pages} {313} (\bibinfo {year} {1992})}\BibitemShut
  {NoStop}%
\bibitem [{\citenamefont {Ross}\ \emph {et~al.}(2021)\citenamefont {Ross},
  \citenamefont {Chastain}, \citenamefont {Vassilenko}, \citenamefont
  {Etemadifar}, \citenamefont {Zimmermann}, \citenamefont {Quesnel},
  \citenamefont {Eert}, \citenamefont {Solomon}, \citenamefont {Patankar},
  \citenamefont {Posacka} \emph {et~al.}}]{ross2021pervasive}%
  \BibitemOpen
  \bibfield  {author} {\bibinfo {author} {\bibfnamefont {P.~S.}\ \bibnamefont
  {Ross}}, \bibinfo {author} {\bibfnamefont {S.}~\bibnamefont {Chastain}},
  \bibinfo {author} {\bibfnamefont {E.}~\bibnamefont {Vassilenko}}, \bibinfo
  {author} {\bibfnamefont {A.}~\bibnamefont {Etemadifar}}, \bibinfo {author}
  {\bibfnamefont {S.}~\bibnamefont {Zimmermann}}, \bibinfo {author}
  {\bibfnamefont {S.-A.}\ \bibnamefont {Quesnel}}, \bibinfo {author}
  {\bibfnamefont {J.}~\bibnamefont {Eert}}, \bibinfo {author} {\bibfnamefont
  {E.}~\bibnamefont {Solomon}}, \bibinfo {author} {\bibfnamefont
  {S.}~\bibnamefont {Patankar}}, \bibinfo {author} {\bibfnamefont {A.~M.}\
  \bibnamefont {Posacka}}, \emph {et~al.},\ }\bibfield  {title} {\bibinfo
  {title} {{Pervasive distribution of polyester fibres in the {A}rctic {O}cean
  is driven by {A}tlantic inputs}},\ }\href@noop {} {\bibfield  {journal}
  {\bibinfo  {journal} {Nat. Commun.}\ }\textbf {\bibinfo {volume} {12}},\
  \bibinfo {pages} {106} (\bibinfo {year} {2021})}\BibitemShut {NoStop}%
\bibitem [{\citenamefont {Ugwu}\ \emph {et~al.}(2021)\citenamefont {Ugwu},
  \citenamefont {Herrera},\ and\ \citenamefont
  {G{\'o}mez}}]{ugwu2021microplastics}%
  \BibitemOpen
  \bibfield  {author} {\bibinfo {author} {\bibfnamefont {K.}~\bibnamefont
  {Ugwu}}, \bibinfo {author} {\bibfnamefont {A.}~\bibnamefont {Herrera}},\ and\
  \bibinfo {author} {\bibfnamefont {M.}~\bibnamefont {G{\'o}mez}},\ }\bibfield
  {title} {\bibinfo {title} {{Microplastics in marine biota: {A} review}},\
  }\href@noop {} {\bibfield  {journal} {\bibinfo  {journal} {Mar. Pollut.
  Bull.}\ }\textbf {\bibinfo {volume} {169}},\ \bibinfo {pages} {112540}
  (\bibinfo {year} {2021})}\BibitemShut {NoStop}%
\bibitem [{\citenamefont {Gruber}\ \emph {et~al.}(2023)\citenamefont {Gruber},
  \citenamefont {Stadlbauer}, \citenamefont {Pichler}, \citenamefont
  {Resch-Fauster}, \citenamefont {Todorovic}, \citenamefont {Meisel},
  \citenamefont {Trawoeger}, \citenamefont {Holl{\'o}czki}, \citenamefont
  {Turner}, \citenamefont {Wadsak} \emph {et~al.}}]{gruber2023waste}%
  \BibitemOpen
  \bibfield  {author} {\bibinfo {author} {\bibfnamefont {E.~S.}\ \bibnamefont
  {Gruber}}, \bibinfo {author} {\bibfnamefont {V.}~\bibnamefont {Stadlbauer}},
  \bibinfo {author} {\bibfnamefont {V.}~\bibnamefont {Pichler}}, \bibinfo
  {author} {\bibfnamefont {K.}~\bibnamefont {Resch-Fauster}}, \bibinfo {author}
  {\bibfnamefont {A.}~\bibnamefont {Todorovic}}, \bibinfo {author}
  {\bibfnamefont {T.~C.}\ \bibnamefont {Meisel}}, \bibinfo {author}
  {\bibfnamefont {S.}~\bibnamefont {Trawoeger}}, \bibinfo {author}
  {\bibfnamefont {O.}~\bibnamefont {Holl{\'o}czki}}, \bibinfo {author}
  {\bibfnamefont {S.~D.}\ \bibnamefont {Turner}}, \bibinfo {author}
  {\bibfnamefont {W.}~\bibnamefont {Wadsak}}, \emph {et~al.},\ }\bibfield
  {title} {\bibinfo {title} {To waste or not to waste: questioning potential
  health risks of micro-and nanoplastics with a focus on their ingestion and
  potential carcinogenicity},\ }\href@noop {} {\bibfield  {journal} {\bibinfo
  {journal} {Expos. Health}\ }\textbf {\bibinfo {volume} {15}},\ \bibinfo
  {pages} {33} (\bibinfo {year} {2023})}\BibitemShut {NoStop}%
\bibitem [{\citenamefont {Leslie}\ \emph {et~al.}(2022)\citenamefont {Leslie},
  \citenamefont {Van~Velzen}, \citenamefont {Brandsma}, \citenamefont
  {Vethaak}, \citenamefont {Garcia-Vallejo},\ and\ \citenamefont
  {Lamoree}}]{leslie2022discovery}%
  \BibitemOpen
  \bibfield  {author} {\bibinfo {author} {\bibfnamefont {H.~A.}\ \bibnamefont
  {Leslie}}, \bibinfo {author} {\bibfnamefont {M.~J.}\ \bibnamefont
  {Van~Velzen}}, \bibinfo {author} {\bibfnamefont {S.~H.}\ \bibnamefont
  {Brandsma}}, \bibinfo {author} {\bibfnamefont {A.~D.}\ \bibnamefont
  {Vethaak}}, \bibinfo {author} {\bibfnamefont {J.~J.}\ \bibnamefont
  {Garcia-Vallejo}},\ and\ \bibinfo {author} {\bibfnamefont {M.~H.}\
  \bibnamefont {Lamoree}},\ }\bibfield  {title} {\bibinfo {title} {Discovery
  and quantification of plastic particle pollution in human blood},\
  }\href@noop {} {\bibfield  {journal} {\bibinfo  {journal} {Environ. Int.}\
  }\textbf {\bibinfo {volume} {163}},\ \bibinfo {pages} {107199} (\bibinfo
  {year} {2022})}\BibitemShut {NoStop}%
\bibitem [{\citenamefont {Chubarenko}\ \emph {et~al.}(2016)\citenamefont
  {Chubarenko}, \citenamefont {Bagaev}, \citenamefont {Zobkov},\ and\
  \citenamefont {Esiukova}}]{chubarenko2016some}%
  \BibitemOpen
  \bibfield  {author} {\bibinfo {author} {\bibfnamefont {I.}~\bibnamefont
  {Chubarenko}}, \bibinfo {author} {\bibfnamefont {A.}~\bibnamefont {Bagaev}},
  \bibinfo {author} {\bibfnamefont {M.}~\bibnamefont {Zobkov}},\ and\ \bibinfo
  {author} {\bibfnamefont {E.}~\bibnamefont {Esiukova}},\ }\bibfield  {title}
  {\bibinfo {title} {On some physical and dynamical properties of microplastic
  particles in marine environment},\ }\href@noop {} {\bibfield  {journal}
  {\bibinfo  {journal} {Mar. Pollut. Bull.}\ }\textbf {\bibinfo {volume}
  {108}},\ \bibinfo {pages} {105} (\bibinfo {year} {2016})}\BibitemShut
  {NoStop}%
\bibitem [{\citenamefont {Clark}\ \emph {et~al.}(2023)\citenamefont {Clark},
  \citenamefont {DiBenedetto}, \citenamefont {Ouellette},\ and\ \citenamefont
  {Koseff}}]{clark2023dispersion}%
  \BibitemOpen
  \bibfield  {author} {\bibinfo {author} {\bibfnamefont {L.~K.}\ \bibnamefont
  {Clark}}, \bibinfo {author} {\bibfnamefont {M.~H.}\ \bibnamefont
  {DiBenedetto}}, \bibinfo {author} {\bibfnamefont {N.~T.}\ \bibnamefont
  {Ouellette}},\ and\ \bibinfo {author} {\bibfnamefont {J.~R.}\ \bibnamefont
  {Koseff}},\ }\bibfield  {title} {\bibinfo {title} {Dispersion of finite-size,
  non-spherical particles by waves and currents},\ }\href@noop {} {\bibfield
  {journal} {\bibinfo  {journal} {J. Fluid Mech.}\ }\textbf {\bibinfo {volume}
  {954}},\ \bibinfo {pages} {A3} (\bibinfo {year} {2023})}\BibitemShut
  {NoStop}%
\bibitem [{\citenamefont {Tatsii}\ \emph {et~al.}(2023)\citenamefont {Tatsii},
  \citenamefont {Bucci}, \citenamefont {Bhowmick}, \citenamefont {Guettler},
  \citenamefont {Bakels}, \citenamefont {Bagheri},\ and\ \citenamefont
  {Stohl}}]{tatsii2023shape}%
  \BibitemOpen
  \bibfield  {author} {\bibinfo {author} {\bibfnamefont {D.}~\bibnamefont
  {Tatsii}}, \bibinfo {author} {\bibfnamefont {S.}~\bibnamefont {Bucci}},
  \bibinfo {author} {\bibfnamefont {T.}~\bibnamefont {Bhowmick}}, \bibinfo
  {author} {\bibfnamefont {J.}~\bibnamefont {Guettler}}, \bibinfo {author}
  {\bibfnamefont {L.}~\bibnamefont {Bakels}}, \bibinfo {author} {\bibfnamefont
  {G.}~\bibnamefont {Bagheri}},\ and\ \bibinfo {author} {\bibfnamefont
  {A.}~\bibnamefont {Stohl}},\ }\bibfield  {title} {\bibinfo {title} {Shape
  matters: long-range transport of microplastic fibers in the atmosphere},\
  }\href@noop {} {\bibfield  {journal} {\bibinfo  {journal} {Environ. Sci.
  Technol.}\ } (\bibinfo {year} {2023})}\BibitemShut {NoStop}%
\bibitem [{\citenamefont {Huntington}\ \emph {et~al.}(2020)\citenamefont
  {Huntington}, \citenamefont {Corcoran}, \citenamefont {Jantunen},
  \citenamefont {Thaysen}, \citenamefont {Bernstein}, \citenamefont {Stern},\
  and\ \citenamefont {Rochman}}]{huntington2020first}%
  \BibitemOpen
  \bibfield  {author} {\bibinfo {author} {\bibfnamefont {A.}~\bibnamefont
  {Huntington}}, \bibinfo {author} {\bibfnamefont {P.~L.}\ \bibnamefont
  {Corcoran}}, \bibinfo {author} {\bibfnamefont {L.}~\bibnamefont {Jantunen}},
  \bibinfo {author} {\bibfnamefont {C.}~\bibnamefont {Thaysen}}, \bibinfo
  {author} {\bibfnamefont {S.}~\bibnamefont {Bernstein}}, \bibinfo {author}
  {\bibfnamefont {G.~A.}\ \bibnamefont {Stern}},\ and\ \bibinfo {author}
  {\bibfnamefont {C.~M.}\ \bibnamefont {Rochman}},\ }\bibfield  {title}
  {\bibinfo {title} {{A first assessment of microplastics and other
  anthropogenic particles in {H}udson {B}ay and the surrounding eastern
  {C}anadian {A}rctic waters of {N}unavut}},\ }\href@noop {} {\bibfield
  {journal} {\bibinfo  {journal} {Facets}\ }\textbf {\bibinfo {volume} {5}},\
  \bibinfo {pages} {432} (\bibinfo {year} {2020})}\BibitemShut {NoStop}%
\bibitem [{\citenamefont {Kooi}\ and\ \citenamefont
  {Koelmans}(2019)}]{kooi2019simplifying}%
  \BibitemOpen
  \bibfield  {author} {\bibinfo {author} {\bibfnamefont {M.}~\bibnamefont
  {Kooi}}\ and\ \bibinfo {author} {\bibfnamefont {A.~A.}\ \bibnamefont
  {Koelmans}},\ }\bibfield  {title} {\bibinfo {title} {Simplifying microplastic
  via continuous probability distributions for size, shape, and density},\
  }\href@noop {} {\bibfield  {journal} {\bibinfo  {journal} {Environ. Sci.
  Technol. Lett.}\ }\textbf {\bibinfo {volume} {6}},\ \bibinfo {pages} {551}
  (\bibinfo {year} {2019})}\BibitemShut {NoStop}%
\bibitem [{\citenamefont {Parsa}\ \emph {et~al.}(2012)\citenamefont {Parsa},
  \citenamefont {Calzavarini}, \citenamefont {Toschi},\ and\ \citenamefont
  {Voth}}]{parsa2012rotation}%
  \BibitemOpen
  \bibfield  {author} {\bibinfo {author} {\bibfnamefont {S.}~\bibnamefont
  {Parsa}}, \bibinfo {author} {\bibfnamefont {E.}~\bibnamefont {Calzavarini}},
  \bibinfo {author} {\bibfnamefont {F.}~\bibnamefont {Toschi}},\ and\ \bibinfo
  {author} {\bibfnamefont {G.~A.}\ \bibnamefont {Voth}},\ }\bibfield  {title}
  {\bibinfo {title} {Rotation rate of rods in turbulent fluid flow},\
  }\href@noop {} {\bibfield  {journal} {\bibinfo  {journal} {Phys. Rev. Lett.}\
  }\textbf {\bibinfo {volume} {109}},\ \bibinfo {pages} {134501} (\bibinfo
  {year} {2012})}\BibitemShut {NoStop}%
\bibitem [{\citenamefont {Brizzolara}\ \emph {et~al.}(2021)\citenamefont
  {Brizzolara}, \citenamefont {Rosti}, \citenamefont {Olivieri}, \citenamefont
  {Brandt}, \citenamefont {Holzner},\ and\ \citenamefont
  {Mazzino}}]{Brizzolara2021}%
  \BibitemOpen
  \bibfield  {author} {\bibinfo {author} {\bibfnamefont {S.}~\bibnamefont
  {Brizzolara}}, \bibinfo {author} {\bibfnamefont {M.~E.}\ \bibnamefont
  {Rosti}}, \bibinfo {author} {\bibfnamefont {S.}~\bibnamefont {Olivieri}},
  \bibinfo {author} {\bibfnamefont {L.}~\bibnamefont {Brandt}}, \bibinfo
  {author} {\bibfnamefont {M.}~\bibnamefont {Holzner}},\ and\ \bibinfo {author}
  {\bibfnamefont {A.}~\bibnamefont {Mazzino}},\ }\bibfield  {title} {\bibinfo
  {title} {Fiber tracking velocimetry for two-point statistics of turbulence},\
  }\href@noop {} {\bibfield  {journal} {\bibinfo  {journal} {Phys. Rev. X}\
  }\textbf {\bibinfo {volume} {11}},\ \bibinfo {pages} {031060} (\bibinfo
  {year} {2021})}\BibitemShut {NoStop}%
\bibitem [{\citenamefont {Byron}\ \emph {et~al.}(2015)\citenamefont {Byron},
  \citenamefont {Einarsson}, \citenamefont {Gustavsson}, \citenamefont {Voth},
  \citenamefont {Mehlig},\ and\ \citenamefont {Variano}}]{byron2015shape}%
  \BibitemOpen
  \bibfield  {author} {\bibinfo {author} {\bibfnamefont {M.}~\bibnamefont
  {Byron}}, \bibinfo {author} {\bibfnamefont {J.}~\bibnamefont {Einarsson}},
  \bibinfo {author} {\bibfnamefont {K.}~\bibnamefont {Gustavsson}}, \bibinfo
  {author} {\bibfnamefont {G.}~\bibnamefont {Voth}}, \bibinfo {author}
  {\bibfnamefont {B.}~\bibnamefont {Mehlig}},\ and\ \bibinfo {author}
  {\bibfnamefont {E.}~\bibnamefont {Variano}},\ }\bibfield  {title} {\bibinfo
  {title} {Shape-dependence of particle rotation in isotropic turbulence},\
  }\href@noop {} {\bibfield  {journal} {\bibinfo  {journal} {Phys. Fluids}\
  }\textbf {\bibinfo {volume} {27}} (\bibinfo {year} {2015})}\BibitemShut
  {NoStop}%
\bibitem [{\citenamefont {Zhao}\ \emph {et~al.}(2015)\citenamefont {Zhao},
  \citenamefont {Challabotla}, \citenamefont {Andersson},\ and\ \citenamefont
  {Variano}}]{zhao2015rotation}%
  \BibitemOpen
  \bibfield  {author} {\bibinfo {author} {\bibfnamefont {L.}~\bibnamefont
  {Zhao}}, \bibinfo {author} {\bibfnamefont {N.~R.}\ \bibnamefont
  {Challabotla}}, \bibinfo {author} {\bibfnamefont {H.~I.}\ \bibnamefont
  {Andersson}},\ and\ \bibinfo {author} {\bibfnamefont {E.~A.}\ \bibnamefont
  {Variano}},\ }\bibfield  {title} {\bibinfo {title} {Rotation of nonspherical
  particles in turbulent channel flow},\ }\href@noop {} {\bibfield  {journal}
  {\bibinfo  {journal} {Phys. Rev. Lett.}\ }\textbf {\bibinfo {volume} {115}},\
  \bibinfo {pages} {244501} (\bibinfo {year} {2015})}\BibitemShut {NoStop}%
\bibitem [{\citenamefont {Marchioli}\ and\ \citenamefont
  {Soldati}(2013)}]{marchioli2013rotation}%
  \BibitemOpen
  \bibfield  {author} {\bibinfo {author} {\bibfnamefont {C.}~\bibnamefont
  {Marchioli}}\ and\ \bibinfo {author} {\bibfnamefont {A.}~\bibnamefont
  {Soldati}},\ }\bibfield  {title} {\bibinfo {title} {Rotation statistics of
  fibers in wall shear turbulence},\ }\href@noop {} {\bibfield  {journal}
  {\bibinfo  {journal} {Acta Mech.}\ }\textbf {\bibinfo {volume} {224}},\
  \bibinfo {pages} {2311} (\bibinfo {year} {2013})}\BibitemShut {NoStop}%
\bibitem [{\citenamefont {Marchioli}\ \emph {et~al.}(2016)\citenamefont
  {Marchioli}, \citenamefont {Zhao},\ and\ \citenamefont
  {Andersson}}]{marchioli2016relative}%
  \BibitemOpen
  \bibfield  {author} {\bibinfo {author} {\bibfnamefont {C.}~\bibnamefont
  {Marchioli}}, \bibinfo {author} {\bibfnamefont {L.}~\bibnamefont {Zhao}},\
  and\ \bibinfo {author} {\bibfnamefont {H.}~\bibnamefont {Andersson}},\
  }\bibfield  {title} {\bibinfo {title} {On the relative rotational motion
  between rigid fibers and fluid in turbulent channel flow},\ }\href@noop {}
  {\bibfield  {journal} {\bibinfo  {journal} {Phys. Fluids}\ }\textbf {\bibinfo
  {volume} {28}} (\bibinfo {year} {2016})}\BibitemShut {NoStop}%
\bibitem [{\citenamefont {Yang}\ \emph {et~al.}(2018)\citenamefont {Yang},
  \citenamefont {Zhao},\ and\ \citenamefont {Andersson}}]{yang2018mean}%
  \BibitemOpen
  \bibfield  {author} {\bibinfo {author} {\bibfnamefont {K.}~\bibnamefont
  {Yang}}, \bibinfo {author} {\bibfnamefont {L.}~\bibnamefont {Zhao}},\ and\
  \bibinfo {author} {\bibfnamefont {H.~I.}\ \bibnamefont {Andersson}},\
  }\bibfield  {title} {\bibinfo {title} {Mean shear versus orientation
  isotropy: effects on inertialess spheroids’ rotation mode in wall
  turbulence},\ }\href@noop {} {\bibfield  {journal} {\bibinfo  {journal} {J.
  Fluid Mech.}\ }\textbf {\bibinfo {volume} {844}},\ \bibinfo {pages} {796}
  (\bibinfo {year} {2018})}\BibitemShut {NoStop}%
\bibitem [{\citenamefont {Baker}\ and\ \citenamefont
  {Coletti}(2022)}]{baker2022experimental}%
  \BibitemOpen
  \bibfield  {author} {\bibinfo {author} {\bibfnamefont {L.~J.}\ \bibnamefont
  {Baker}}\ and\ \bibinfo {author} {\bibfnamefont {F.}~\bibnamefont
  {Coletti}},\ }\bibfield  {title} {\bibinfo {title} {Experimental
  investigation of inertial fibres and disks in a turbulent boundary layer},\
  }\href@noop {} {\bibfield  {journal} {\bibinfo  {journal} {J. Fluid Mech.}\
  }\textbf {\bibinfo {volume} {943}},\ \bibinfo {pages} {A27} (\bibinfo {year}
  {2022})}\BibitemShut {NoStop}%
\bibitem [{\citenamefont {Shaik}\ and\ \citenamefont {van
  Hout}(2023)}]{shaik2023kinematics}%
  \BibitemOpen
  \bibfield  {author} {\bibinfo {author} {\bibfnamefont {S.}~\bibnamefont
  {Shaik}}\ and\ \bibinfo {author} {\bibfnamefont {R.}~\bibnamefont {van
  Hout}},\ }\bibfield  {title} {\bibinfo {title} {Kinematics of rigid fibers in
  a turbulent channel flow},\ }\href@noop {} {\bibfield  {journal} {\bibinfo
  {journal} {Int. J. Multiphas. Flow}\ }\textbf {\bibinfo {volume} {158}},\
  \bibinfo {pages} {104262} (\bibinfo {year} {2023})}\BibitemShut {NoStop}%
\bibitem [{\citenamefont {Hoseini}\ \emph {et~al.}(2015)\citenamefont
  {Hoseini}, \citenamefont {Lundell},\ and\ \citenamefont
  {Andersson}}]{hoseini2015finite}%
  \BibitemOpen
  \bibfield  {author} {\bibinfo {author} {\bibfnamefont {A.~A.}\ \bibnamefont
  {Hoseini}}, \bibinfo {author} {\bibfnamefont {F.}~\bibnamefont {Lundell}},\
  and\ \bibinfo {author} {\bibfnamefont {H.~I.}\ \bibnamefont {Andersson}},\
  }\bibfield  {title} {\bibinfo {title} {Finite-length effects on dynamical
  behavior of rod-like particles in wall-bounded turbulent flow},\ }\href@noop
  {} {\bibfield  {journal} {\bibinfo  {journal} {Int. J. Multiphas. Flow}\
  }\textbf {\bibinfo {volume} {76}},\ \bibinfo {pages} {13} (\bibinfo {year}
  {2015})}\BibitemShut {NoStop}%
\bibitem [{\citenamefont {Giurgiu}\ \emph {et~al.}(2023)\citenamefont
  {Giurgiu}, \citenamefont {Caridi}, \citenamefont {Alipour}, \citenamefont
  {De~Paoli},\ and\ \citenamefont {Soldati}}]{giurgiu2023tu}%
  \BibitemOpen
  \bibfield  {author} {\bibinfo {author} {\bibfnamefont {V.}~\bibnamefont
  {Giurgiu}}, \bibinfo {author} {\bibfnamefont {G.~C.~A.}\ \bibnamefont
  {Caridi}}, \bibinfo {author} {\bibfnamefont {M.}~\bibnamefont {Alipour}},
  \bibinfo {author} {\bibfnamefont {M.}~\bibnamefont {De~Paoli}},\ and\
  \bibinfo {author} {\bibfnamefont {A.}~\bibnamefont {Soldati}},\ }\bibfield
  {title} {\bibinfo {title} {{The {T}U {W}ien {T}urbulent {W}ater {C}hannel:
  {F}low control loop and three-dimensional reconstruction of anisotropic
  particle dynamics}},\ }\href@noop {} {\bibfield  {journal} {\bibinfo
  {journal} {Rev. Sci. Instrum.}\ }\textbf {\bibinfo {volume} {94}} (\bibinfo
  {year} {2023})}\BibitemShut {NoStop}%
\bibitem [{Sup(2024{\natexlab{a}})}]{SupplMat_exp_det}%
  \BibitemOpen
  \href@noop {} {\bibinfo {title} {Supplemental material for additional details
  on methodology, which includes {R}efs.
  \cite{kind2013vdi,elsinga2006tomographic,jiang2024priv,jiang2020rotation}}}
  (\bibinfo {year} {2024}{\natexlab{a}})\BibitemShut {NoStop}%
\bibitem [{\citenamefont {Moser}\ \emph {et~al.}(1999)\citenamefont {Moser},
  \citenamefont {Kim},\ and\ \citenamefont {Mansour}}]{moser1999direct}%
  \BibitemOpen
  \bibfield  {author} {\bibinfo {author} {\bibfnamefont {R.~D.}\ \bibnamefont
  {Moser}}, \bibinfo {author} {\bibfnamefont {J.}~\bibnamefont {Kim}},\ and\
  \bibinfo {author} {\bibfnamefont {N.~N.}\ \bibnamefont {Mansour}},\
  }\bibfield  {title} {\bibinfo {title} {{Direct numerical simulation of
  turbulent channel flow up to $Re_\tau=590$}},\ }\href@noop {} {\bibfield
  {journal} {\bibinfo  {journal} {Phys. Fluids}\ }\textbf {\bibinfo {volume}
  {11}},\ \bibinfo {pages} {943} (\bibinfo {year} {1999})}\BibitemShut
  {NoStop}%
\bibitem [{\citenamefont {Del~Alamo}\ \emph {et~al.}(2004)\citenamefont
  {Del~Alamo}, \citenamefont {Jim{\'e}nez}, \citenamefont {Zandonade},\ and\
  \citenamefont {Moser}}]{del2004scaling}%
  \BibitemOpen
  \bibfield  {author} {\bibinfo {author} {\bibfnamefont {J.~C.}\ \bibnamefont
  {Del~Alamo}}, \bibinfo {author} {\bibfnamefont {J.}~\bibnamefont
  {Jim{\'e}nez}}, \bibinfo {author} {\bibfnamefont {P.}~\bibnamefont
  {Zandonade}},\ and\ \bibinfo {author} {\bibfnamefont {R.~D.}\ \bibnamefont
  {Moser}},\ }\bibfield  {title} {\bibinfo {title} {Scaling of the energy
  spectra of turbulent channels},\ }\href@noop {} {\bibfield  {journal}
  {\bibinfo  {journal} {J. Fluid Mech.}\ }\textbf {\bibinfo {volume} {500}},\
  \bibinfo {pages} {135} (\bibinfo {year} {2004})}\BibitemShut {NoStop}%
\bibitem [{Sup(2024{\natexlab{b}})}]{SupplMat_settling}%
  \BibitemOpen
  \href@noop {} {\bibinfo {title} {Supplemental material with details about the
  settling behaviour of fibers due to the density mismatch with the working
  fluid}} (\bibinfo {year} {2024}{\natexlab{b}})\BibitemShut {NoStop}%
\bibitem [{\citenamefont {Bhowmick}\ \emph {et~al.}(2024)\citenamefont
  {Bhowmick}, \citenamefont {Seesing}, \citenamefont {Gustavsson},
  \citenamefont {Guettler}, \citenamefont {Wang}, \citenamefont {Pumir},
  \citenamefont {Mehlig},\ and\ \citenamefont {Bagheri}}]{bhowmick2024inertia}%
  \BibitemOpen
  \bibfield  {author} {\bibinfo {author} {\bibfnamefont {T.}~\bibnamefont
  {Bhowmick}}, \bibinfo {author} {\bibfnamefont {J.}~\bibnamefont {Seesing}},
  \bibinfo {author} {\bibfnamefont {K.}~\bibnamefont {Gustavsson}}, \bibinfo
  {author} {\bibfnamefont {J.}~\bibnamefont {Guettler}}, \bibinfo {author}
  {\bibfnamefont {Y.}~\bibnamefont {Wang}}, \bibinfo {author} {\bibfnamefont
  {A.}~\bibnamefont {Pumir}}, \bibinfo {author} {\bibfnamefont
  {B.}~\bibnamefont {Mehlig}},\ and\ \bibinfo {author} {\bibfnamefont
  {G.}~\bibnamefont {Bagheri}},\ }\bibfield  {title} {\bibinfo {title} {Inertia
  induces strong orientation fluctuations of nonspherical atmospheric
  particles},\ }\href@noop {} {\bibfield  {journal} {\bibinfo  {journal} {Phys.
  Rev. Lett.}\ }\textbf {\bibinfo {volume} {132}},\ \bibinfo {pages} {034101}
  (\bibinfo {year} {2024})}\BibitemShut {NoStop}%
\bibitem [{\citenamefont {{The Flocking Shop}}(2022)}]{flock}%
  \BibitemOpen
  \bibfield  {author} {\bibinfo {author} {\bibnamefont {{The Flocking Shop}}},\
  }\href {https://www.theflockingshop.co.uk} {\bibinfo {title} {{{D}{C}{A}
  {F}lock, https://www.theflockingshop.co.uk}}} (\bibinfo {year}
  {2022})\BibitemShut {NoStop}%
\bibitem [{\citenamefont {Bernstein}\ and\ \citenamefont
  {Shapiro}(1994)}]{bernstein1994direct}%
  \BibitemOpen
  \bibfield  {author} {\bibinfo {author} {\bibfnamefont {O.}~\bibnamefont
  {Bernstein}}\ and\ \bibinfo {author} {\bibfnamefont {M.}~\bibnamefont
  {Shapiro}},\ }\bibfield  {title} {\bibinfo {title} {Direct determination of
  the orientation distribution function of cylindrical particles immersed in
  laminar and turbulent shear flows},\ }\href@noop {} {\bibfield  {journal}
  {\bibinfo  {journal} {Journal of Aerosol Science}\ }\textbf {\bibinfo
  {volume} {25}},\ \bibinfo {pages} {113} (\bibinfo {year} {1994})}\BibitemShut
  {NoStop}%
\bibitem [{\citenamefont {Alipour}\ \emph {et~al.}(2022)\citenamefont
  {Alipour}, \citenamefont {De~Paoli},\ and\ \citenamefont
  {Soldati}}]{alipour2022influence}%
  \BibitemOpen
  \bibfield  {author} {\bibinfo {author} {\bibfnamefont {M.}~\bibnamefont
  {Alipour}}, \bibinfo {author} {\bibfnamefont {M.}~\bibnamefont {De~Paoli}},\
  and\ \bibinfo {author} {\bibfnamefont {A.}~\bibnamefont {Soldati}},\
  }\bibfield  {title} {\bibinfo {title} {{Influence of {R}eynolds number on the
  dynamics of rigid, slender and non-axisymmetric fibres in channel flow
  turbulence}},\ }\href@noop {} {\bibfield  {journal} {\bibinfo  {journal} {J.
  Fluid Mech.}\ }\textbf {\bibinfo {volume} {934}},\ \bibinfo {pages} {A18}
  (\bibinfo {year} {2022})}\BibitemShut {NoStop}%
\bibitem [{\citenamefont {Bounoua}\ \emph {et~al.}(2018)\citenamefont
  {Bounoua}, \citenamefont {Bouchet},\ and\ \citenamefont
  {Verhille}}]{bounoua2018tumbling}%
  \BibitemOpen
  \bibfield  {author} {\bibinfo {author} {\bibfnamefont {S.}~\bibnamefont
  {Bounoua}}, \bibinfo {author} {\bibfnamefont {G.}~\bibnamefont {Bouchet}},\
  and\ \bibinfo {author} {\bibfnamefont {G.}~\bibnamefont {Verhille}},\
  }\bibfield  {title} {\bibinfo {title} {Tumbling of inertial fibers in
  turbulence},\ }\href@noop {} {\bibfield  {journal} {\bibinfo  {journal}
  {Phys. Rev. Lett.}\ }\textbf {\bibinfo {volume} {121}},\ \bibinfo {pages}
  {124502} (\bibinfo {year} {2018})}\BibitemShut {NoStop}%
\bibitem [{\citenamefont {Alipour}\ \emph {et~al.}(2021)\citenamefont
  {Alipour}, \citenamefont {De~Paoli}, \citenamefont {Ghaemi},\ and\
  \citenamefont {Soldati}}]{alipour2021long}%
  \BibitemOpen
  \bibfield  {author} {\bibinfo {author} {\bibfnamefont {M.}~\bibnamefont
  {Alipour}}, \bibinfo {author} {\bibfnamefont {M.}~\bibnamefont {De~Paoli}},
  \bibinfo {author} {\bibfnamefont {S.}~\bibnamefont {Ghaemi}},\ and\ \bibinfo
  {author} {\bibfnamefont {A.}~\bibnamefont {Soldati}},\ }\bibfield  {title}
  {\bibinfo {title} {Long non-axisymmetric fibres in turbulent channel flow},\
  }\href@noop {} {\bibfield  {journal} {\bibinfo  {journal} {J. Fluid Mech.}\
  }\textbf {\bibinfo {volume} {916}},\ \bibinfo {pages} {A3} (\bibinfo {year}
  {2021})}\BibitemShut {NoStop}%
\bibitem [{Sup(2024{\natexlab{c}})}]{SupplMat_method}%
  \BibitemOpen
  \href@noop {} {\bibinfo {title} {Supplemental material with details about the
  fiber measurement volume, resolution, acquisition frequency and sample size}}
  (\bibinfo {year} {2024}{\natexlab{c}})\BibitemShut {NoStop}%
\bibitem [{Sup(2024{\natexlab{d}})}]{SupplMat_Fig_curv_pdf}%
  \BibitemOpen
  \href@noop {} {\bibinfo {title} {Supplemental material for p.d.f. of
  dimensionless curvature}} (\bibinfo {year} {2024}{\natexlab{d}})\BibitemShut
  {NoStop}%
\bibitem [{Sup(2024{\natexlab{e}})}]{SupplMat_rotation_rates_comp}%
  \BibitemOpen
  \href@noop {} {\bibinfo {title} {Supplemental material for the computational
  method of rotation rates, which includes {R}efs.
  \cite{cleveland1979robust,lynch2017modern}}} (\bibinfo {year}
  {2024}{\natexlab{e}})\BibitemShut {NoStop}%
\bibitem [{\citenamefont {Parsa}\ and\ \citenamefont
  {Voth}(2014)}]{parsa2014inertial}%
  \BibitemOpen
  \bibfield  {author} {\bibinfo {author} {\bibfnamefont {S.}~\bibnamefont
  {Parsa}}\ and\ \bibinfo {author} {\bibfnamefont {G.~A.}\ \bibnamefont
  {Voth}},\ }\bibfield  {title} {\bibinfo {title} {Inertial range scaling in
  rotations of long rods in turbulence},\ }\href@noop {} {\bibfield  {journal}
  {\bibinfo  {journal} {Phys. Rev. Lett.}\ }\textbf {\bibinfo {volume} {112}},\
  \bibinfo {pages} {024501} (\bibinfo {year} {2014})}\BibitemShut {NoStop}%
\bibitem [{Sup(2024{\natexlab{f}})}]{SupplMat_Fig_tumbl_pdf_number_sample}%
  \BibitemOpen
  \href@noop {} {\bibinfo {title} {Supplemental material the effect of sample
  size on the normalized squared tumbling rate}} (\bibinfo {year}
  {2024}{\natexlab{f}})\BibitemShut {NoStop}%
\bibitem [{Sup(2024{\natexlab{g}})}]{SupplMat_Fig_curv_effect}%
  \BibitemOpen
  \href@noop {} {\bibinfo {title} {Supplemental material the effect of the
  dimensionless curvature on mean squared tumbling and spinning rates}}
  (\bibinfo {year} {2024}{\natexlab{g}})\BibitemShut {NoStop}%
\bibitem [{\citenamefont {Oehmke}\ \emph {et~al.}(2021)\citenamefont {Oehmke},
  \citenamefont {Bordoloi}, \citenamefont {Variano},\ and\ \citenamefont
  {Verhille}}]{oehmke2021spinning}%
  \BibitemOpen
  \bibfield  {author} {\bibinfo {author} {\bibfnamefont {T.~B.}\ \bibnamefont
  {Oehmke}}, \bibinfo {author} {\bibfnamefont {A.~D.}\ \bibnamefont
  {Bordoloi}}, \bibinfo {author} {\bibfnamefont {E.}~\bibnamefont {Variano}},\
  and\ \bibinfo {author} {\bibfnamefont {G.}~\bibnamefont {Verhille}},\
  }\bibfield  {title} {\bibinfo {title} {Spinning and tumbling of long fibers
  in isotropic turbulence},\ }\href@noop {} {\bibfield  {journal} {\bibinfo
  {journal} {Phys. Rev. Fluids}\ }\textbf {\bibinfo {volume} {6}},\ \bibinfo
  {pages} {044610} (\bibinfo {year} {2021})}\BibitemShut {NoStop}%
\bibitem [{\citenamefont {Shaik}\ \emph {et~al.}(2020)\citenamefont {Shaik},
  \citenamefont {Kuperman}, \citenamefont {Rinsky},\ and\ \citenamefont {van
  Hout}}]{shaik2020measurements}%
  \BibitemOpen
  \bibfield  {author} {\bibinfo {author} {\bibfnamefont {S.}~\bibnamefont
  {Shaik}}, \bibinfo {author} {\bibfnamefont {S.}~\bibnamefont {Kuperman}},
  \bibinfo {author} {\bibfnamefont {V.}~\bibnamefont {Rinsky}},\ and\ \bibinfo
  {author} {\bibfnamefont {R.}~\bibnamefont {van Hout}},\ }\bibfield  {title}
  {\bibinfo {title} {Measurements of length effects on the dynamics of rigid
  fibers in a turbulent channel flow},\ }\href@noop {} {\bibfield  {journal}
  {\bibinfo  {journal} {Phys. Rev. Fluids}\ }\textbf {\bibinfo {volume} {5}},\
  \bibinfo {pages} {114309} (\bibinfo {year} {2020})}\BibitemShut {NoStop}%
\bibitem [{\citenamefont {Mortensen}\ \emph {et~al.}(2008)\citenamefont
  {Mortensen}, \citenamefont {Andersson}, \citenamefont {Gillissen},\ and\
  \citenamefont {Boersma}}]{mortensen2008dynamics}%
  \BibitemOpen
  \bibfield  {author} {\bibinfo {author} {\bibfnamefont {P.}~\bibnamefont
  {Mortensen}}, \bibinfo {author} {\bibfnamefont {H.}~\bibnamefont
  {Andersson}}, \bibinfo {author} {\bibfnamefont {J.}~\bibnamefont
  {Gillissen}},\ and\ \bibinfo {author} {\bibfnamefont {B.}~\bibnamefont
  {Boersma}},\ }\bibfield  {title} {\bibinfo {title} {Dynamics of prolate
  ellipsoidal particles in a turbulent channel flow},\ }\href@noop {}
  {\bibfield  {journal} {\bibinfo  {journal} {Phys. Fluids}\ }\textbf {\bibinfo
  {volume} {20}} (\bibinfo {year} {2008})}\BibitemShut {NoStop}%
\bibitem [{Sup(2024{\natexlab{h}})}]{SupplMat_Fig_convergence}%
  \BibitemOpen
  \href@noop {} {\bibinfo {title} {Supplemental material for convergence of
  tumbling and spinning rates at three wall-normal positions}} (\bibinfo {year}
  {2024}{\natexlab{h}})\BibitemShut {NoStop}%
\bibitem [{\citenamefont {Marcus}\ \emph {et~al.}(2014)\citenamefont {Marcus},
  \citenamefont {Parsa}, \citenamefont {Kramel}, \citenamefont {Ni},\ and\
  \citenamefont {Voth}}]{marcus2014measurements}%
  \BibitemOpen
  \bibfield  {author} {\bibinfo {author} {\bibfnamefont {G.~G.}\ \bibnamefont
  {Marcus}}, \bibinfo {author} {\bibfnamefont {S.}~\bibnamefont {Parsa}},
  \bibinfo {author} {\bibfnamefont {S.}~\bibnamefont {Kramel}}, \bibinfo
  {author} {\bibfnamefont {R.}~\bibnamefont {Ni}},\ and\ \bibinfo {author}
  {\bibfnamefont {G.~A.}\ \bibnamefont {Voth}},\ }\bibfield  {title} {\bibinfo
  {title} {Measurements of the solid-body rotation of anisotropic particles in
  3{D} turbulence},\ }\href@noop {} {\bibfield  {journal} {\bibinfo  {journal}
  {New J Phys.}\ }\textbf {\bibinfo {volume} {16}},\ \bibinfo {pages} {102001}
  (\bibinfo {year} {2014})}\BibitemShut {NoStop}%
\bibitem [{\citenamefont {Pujara}\ \emph {et~al.}(2021)\citenamefont {Pujara},
  \citenamefont {Arguedas-Leiva}, \citenamefont {Lalescu}, \citenamefont
  {Bramas},\ and\ \citenamefont {Wilczek}}]{pujara2021shape}%
  \BibitemOpen
  \bibfield  {author} {\bibinfo {author} {\bibfnamefont {N.}~\bibnamefont
  {Pujara}}, \bibinfo {author} {\bibfnamefont {J.-A.}\ \bibnamefont
  {Arguedas-Leiva}}, \bibinfo {author} {\bibfnamefont {C.~C.}\ \bibnamefont
  {Lalescu}}, \bibinfo {author} {\bibfnamefont {B.}~\bibnamefont {Bramas}},\
  and\ \bibinfo {author} {\bibfnamefont {M.}~\bibnamefont {Wilczek}},\
  }\bibfield  {title} {\bibinfo {title} {Shape-and scale-dependent coupling
  between spheroids and velocity gradients in turbulence},\ }\href@noop {}
  {\bibfield  {journal} {\bibinfo  {journal} {J. Fluid Mech.}\ }\textbf
  {\bibinfo {volume} {922}},\ \bibinfo {pages} {R6} (\bibinfo {year}
  {2021})}\BibitemShut {NoStop}%
\bibitem [{\citenamefont {Douady}\ \emph {et~al.}(1991)\citenamefont {Douady},
  \citenamefont {Couder},\ and\ \citenamefont {Brachet}}]{douady1991direct}%
  \BibitemOpen
  \bibfield  {author} {\bibinfo {author} {\bibfnamefont {S.}~\bibnamefont
  {Douady}}, \bibinfo {author} {\bibfnamefont {Y.}~\bibnamefont {Couder}},\
  and\ \bibinfo {author} {\bibfnamefont {M.~E.}\ \bibnamefont {Brachet}},\
  }\bibfield  {title} {\bibinfo {title} {Direct observation of the
  intermittency of intense vorticity filaments in turbulence},\ }\href@noop {}
  {\bibfield  {journal} {\bibinfo  {journal} {Phys. Rev. Lett.}\ }\textbf
  {\bibinfo {volume} {67}},\ \bibinfo {pages} {983} (\bibinfo {year}
  {1991})}\BibitemShut {NoStop}%
\bibitem [{\citenamefont {Jimenez}\ and\ \citenamefont
  {Wray}(1998)}]{jimenez1998characteristics}%
  \BibitemOpen
  \bibfield  {author} {\bibinfo {author} {\bibfnamefont {J.}~\bibnamefont
  {Jimenez}}\ and\ \bibinfo {author} {\bibfnamefont {A.~A.}\ \bibnamefont
  {Wray}},\ }\bibfield  {title} {\bibinfo {title} {On the characteristics of
  vortex filaments in isotropic turbulence},\ }\href@noop {} {\bibfield
  {journal} {\bibinfo  {journal} {J. Fluid Mech.}\ }\textbf {\bibinfo {volume}
  {373}},\ \bibinfo {pages} {255} (\bibinfo {year} {1998})}\BibitemShut
  {NoStop}%
\bibitem [{\citenamefont {Robinson}(1991)}]{robinson1991coherent}%
  \BibitemOpen
  \bibfield  {author} {\bibinfo {author} {\bibfnamefont {S.~K.}\ \bibnamefont
  {Robinson}},\ }\bibfield  {title} {\bibinfo {title} {Coherent motions in the
  turbulent boundary layer},\ }\href@noop {} {\bibfield  {journal} {\bibinfo
  {journal} {Annu. Rev. Fluid Mech.}\ }\textbf {\bibinfo {volume} {23}},\
  \bibinfo {pages} {601} (\bibinfo {year} {1991})}\BibitemShut {NoStop}%
\bibitem [{\citenamefont {Adrian}(2007)}]{adrian2007hairpin}%
  \BibitemOpen
  \bibfield  {author} {\bibinfo {author} {\bibfnamefont {R.~J.}\ \bibnamefont
  {Adrian}},\ }\bibfield  {title} {\bibinfo {title} {Hairpin vortex
  organization in wall turbulence},\ }\href@noop {} {\bibfield  {journal}
  {\bibinfo  {journal} {{Phys. Fluids}}\ }\textbf {\bibinfo {volume} {19}}
  (\bibinfo {year} {2007})}\BibitemShut {NoStop}%
\bibitem [{\citenamefont {Yang}\ and\ \citenamefont
  {Pullin}(2011)}]{yang2011geometric}%
  \BibitemOpen
  \bibfield  {author} {\bibinfo {author} {\bibfnamefont {Y.}~\bibnamefont
  {Yang}}\ and\ \bibinfo {author} {\bibfnamefont {D.}~\bibnamefont {Pullin}},\
  }\bibfield  {title} {\bibinfo {title} {Geometric study of {L}agrangian and
  {E}ulerian structures in turbulent channel flow},\ }\href@noop {} {\bibfield
  {journal} {\bibinfo  {journal} {J. Fluid Mech.}\ }\textbf {\bibinfo {volume}
  {674}},\ \bibinfo {pages} {67} (\bibinfo {year} {2011})}\BibitemShut
  {NoStop}%
\bibitem [{\citenamefont {Green}\ \emph {et~al.}(2007)\citenamefont {Green},
  \citenamefont {Rowley},\ and\ \citenamefont {Haller}}]{green2007detection}%
  \BibitemOpen
  \bibfield  {author} {\bibinfo {author} {\bibfnamefont {M.~A.}\ \bibnamefont
  {Green}}, \bibinfo {author} {\bibfnamefont {C.~W.}\ \bibnamefont {Rowley}},\
  and\ \bibinfo {author} {\bibfnamefont {G.}~\bibnamefont {Haller}},\
  }\bibfield  {title} {\bibinfo {title} {Detection of lagrangian coherent
  structures in three-dimensional turbulence},\ }\href@noop {} {\bibfield
  {journal} {\bibinfo  {journal} {J. Fluid Mech.}\ }\textbf {\bibinfo {volume}
  {572}},\ \bibinfo {pages} {111} (\bibinfo {year} {2007})}\BibitemShut
  {NoStop}%
\bibitem [{\citenamefont {Jim{\'e}nez}(2018)}]{jimenez2018coherent}%
  \BibitemOpen
  \bibfield  {author} {\bibinfo {author} {\bibfnamefont {J.}~\bibnamefont
  {Jim{\'e}nez}},\ }\bibfield  {title} {\bibinfo {title} {Coherent structures
  in wall-bounded turbulence},\ }\href@noop {} {\bibfield  {journal} {\bibinfo
  {journal} {J. Fluid Mech.}\ }\textbf {\bibinfo {volume} {842}},\ \bibinfo
  {pages} {P1} (\bibinfo {year} {2018})}\BibitemShut {NoStop}%
\bibitem [{\citenamefont {Sharma}\ and\ \citenamefont
  {McKeon}(2013)}]{sharma2013coherent}%
  \BibitemOpen
  \bibfield  {author} {\bibinfo {author} {\bibfnamefont {A.}~\bibnamefont
  {Sharma}}\ and\ \bibinfo {author} {\bibfnamefont {B.~J.}\ \bibnamefont
  {McKeon}},\ }\bibfield  {title} {\bibinfo {title} {On coherent structure in
  wall turbulence},\ }\href@noop {} {\bibfield  {journal} {\bibinfo  {journal}
  {J. Fluid Mech.}\ }\textbf {\bibinfo {volume} {728}},\ \bibinfo {pages} {196}
  (\bibinfo {year} {2013})}\BibitemShut {NoStop}%
\bibitem [{\citenamefont {Challabotla}\ \emph {et~al.}(2015)\citenamefont
  {Challabotla}, \citenamefont {Zhao},\ and\ \citenamefont
  {Andersson}}]{challabotla2015shape}%
  \BibitemOpen
  \bibfield  {author} {\bibinfo {author} {\bibfnamefont {N.~R.}\ \bibnamefont
  {Challabotla}}, \bibinfo {author} {\bibfnamefont {L.}~\bibnamefont {Zhao}},\
  and\ \bibinfo {author} {\bibfnamefont {H.~I.}\ \bibnamefont {Andersson}},\
  }\bibfield  {title} {\bibinfo {title} {Shape effects on dynamics of
  inertia-free spheroids in wall turbulence},\ }\href@noop {} {\bibfield
  {journal} {\bibinfo  {journal} {Phys. Fluids}\ }\textbf {\bibinfo {volume}
  {27}} (\bibinfo {year} {2015})}\BibitemShut {NoStop}%
\bibitem [{\citenamefont {Zhao}\ and\ \citenamefont
  {Andersson}(2016)}]{zhao2016spheroids}%
  \BibitemOpen
  \bibfield  {author} {\bibinfo {author} {\bibfnamefont {L.}~\bibnamefont
  {Zhao}}\ and\ \bibinfo {author} {\bibfnamefont {H.~I.}\ \bibnamefont
  {Andersson}},\ }\bibfield  {title} {\bibinfo {title} {Why spheroids orient
  preferentially in near-wall turbulence},\ }\href@noop {} {\bibfield
  {journal} {\bibinfo  {journal} {J. Fluid Mech.}\ }\textbf {\bibinfo {volume}
  {807}},\ \bibinfo {pages} {221} (\bibinfo {year} {2016})}\BibitemShut
  {NoStop}%
\bibitem [{\citenamefont {Do-Quang}\ \emph {et~al.}(2014)\citenamefont
  {Do-Quang}, \citenamefont {Amberg}, \citenamefont {Brethouwer},\ and\
  \citenamefont {Johansson}}]{do2014simulation}%
  \BibitemOpen
  \bibfield  {author} {\bibinfo {author} {\bibfnamefont {M.}~\bibnamefont
  {Do-Quang}}, \bibinfo {author} {\bibfnamefont {G.}~\bibnamefont {Amberg}},
  \bibinfo {author} {\bibfnamefont {G.}~\bibnamefont {Brethouwer}},\ and\
  \bibinfo {author} {\bibfnamefont {A.~V.}\ \bibnamefont {Johansson}},\
  }\bibfield  {title} {\bibinfo {title} {Simulation of finite-size fibers in
  turbulent channel flows},\ }\href@noop {} {\bibfield  {journal} {\bibinfo
  {journal} {Phys. Rev. E}\ }\textbf {\bibinfo {volume} {89}},\ \bibinfo
  {pages} {013006} (\bibinfo {year} {2014})}\BibitemShut {NoStop}%
\bibitem [{\citenamefont {Cui}\ \emph {et~al.}(2021)\citenamefont {Cui},
  \citenamefont {Huang}, \citenamefont {Xu}, \citenamefont {Andersson},\ and\
  \citenamefont {Zhao}}]{cui2021alignment}%
  \BibitemOpen
  \bibfield  {author} {\bibinfo {author} {\bibfnamefont {Z.}~\bibnamefont
  {Cui}}, \bibinfo {author} {\bibfnamefont {W.-X.}\ \bibnamefont {Huang}},
  \bibinfo {author} {\bibfnamefont {C.-X.}\ \bibnamefont {Xu}}, \bibinfo
  {author} {\bibfnamefont {H.~I.}\ \bibnamefont {Andersson}},\ and\ \bibinfo
  {author} {\bibfnamefont {L.}~\bibnamefont {Zhao}},\ }\bibfield  {title}
  {\bibinfo {title} {Alignment of slender fibers and thin disks induced by
  coherent structures of wall turbulence},\ }\href@noop {} {\bibfield
  {journal} {\bibinfo  {journal} {Int. J. Multiphas. Flow}\ }\textbf {\bibinfo
  {volume} {145}},\ \bibinfo {pages} {103837} (\bibinfo {year}
  {2021})}\BibitemShut {NoStop}%
\bibitem [{\citenamefont {Ni}\ \emph {et~al.}(2014)\citenamefont {Ni},
  \citenamefont {Ouellette},\ and\ \citenamefont {Voth}}]{ni2014alignment}%
  \BibitemOpen
  \bibfield  {author} {\bibinfo {author} {\bibfnamefont {R.}~\bibnamefont
  {Ni}}, \bibinfo {author} {\bibfnamefont {N.~T.}\ \bibnamefont {Ouellette}},\
  and\ \bibinfo {author} {\bibfnamefont {G.~A.}\ \bibnamefont {Voth}},\
  }\bibfield  {title} {\bibinfo {title} {{Alignment of vorticity and rods with
  {L}agrangian fluid stretching in turbulence}},\ }\href@noop {} {\bibfield
  {journal} {\bibinfo  {journal} {J. Fluid Mech.}\ }\textbf {\bibinfo {volume}
  {743}},\ \bibinfo {pages} {R3} (\bibinfo {year} {2014})}\BibitemShut
  {NoStop}%
\bibitem [{Sup(2024{\natexlab{i}})}]{SupplMat_Fig_curv_effect_on_orientation}%
  \BibitemOpen
  \href@noop {} {\bibinfo {title} {Supplemental material the effect of the
  dimensionless curvature on mean squared tumbling and spinning rates}}
  (\bibinfo {year} {2024}{\natexlab{i}})\BibitemShut {NoStop}%
\bibitem [{\citenamefont {Iwamoto}\ \emph {et~al.}(2002)\citenamefont
  {Iwamoto}, \citenamefont {Suzuki},\ and\ \citenamefont
  {Kasagi}}]{iwamoto2002reynolds}%
  \BibitemOpen
  \bibfield  {author} {\bibinfo {author} {\bibfnamefont {K.}~\bibnamefont
  {Iwamoto}}, \bibinfo {author} {\bibfnamefont {Y.}~\bibnamefont {Suzuki}},\
  and\ \bibinfo {author} {\bibfnamefont {N.}~\bibnamefont {Kasagi}},\
  }\bibfield  {title} {\bibinfo {title} {Reynolds number effect on wall
  turbulence: toward effective feedback control},\ }\href@noop {} {\bibfield
  {journal} {\bibinfo  {journal} {Int. J. Heat Fluid Flow}\ }\textbf {\bibinfo
  {volume} {23}},\ \bibinfo {pages} {678} (\bibinfo {year} {2002})}\BibitemShut
  {NoStop}%
\bibitem [{\citenamefont {Zhao}\ \emph {et~al.}(2019)\citenamefont {Zhao},
  \citenamefont {Challabotla}, \citenamefont {Andersson},\ and\ \citenamefont
  {Variano}}]{zhao2019mapping}%
  \BibitemOpen
  \bibfield  {author} {\bibinfo {author} {\bibfnamefont {L.}~\bibnamefont
  {Zhao}}, \bibinfo {author} {\bibfnamefont {N.~R.}\ \bibnamefont
  {Challabotla}}, \bibinfo {author} {\bibfnamefont {H.~I.}\ \bibnamefont
  {Andersson}},\ and\ \bibinfo {author} {\bibfnamefont {E.~A.}\ \bibnamefont
  {Variano}},\ }\bibfield  {title} {\bibinfo {title} {Mapping spheroid rotation
  modes in turbulent channel flow: effects of shear, turbulence and particle
  inertia},\ }\href@noop {} {\bibfield  {journal} {\bibinfo  {journal} {J.
  Fluid Mech.}\ }\textbf {\bibinfo {volume} {876}},\ \bibinfo {pages} {19}
  (\bibinfo {year} {2019})}\BibitemShut {NoStop}%
\bibitem [{\citenamefont {Kind}\ and\ \citenamefont
  {Martin}(2013)}]{kind2013vdi}%
  \BibitemOpen
  \bibfield  {author} {\bibinfo {author} {\bibfnamefont {M.}~\bibnamefont
  {Kind}}\ and\ \bibinfo {author} {\bibfnamefont {H.}~\bibnamefont {Martin}},\
  }\href@noop {} {\bibinfo {title} {V{D}{I}-{W}{\"a}rmeatlas}} (\bibinfo {year}
  {2013})\BibitemShut {NoStop}%
\bibitem [{\citenamefont {Elsinga}\ \emph {et~al.}(2006)\citenamefont
  {Elsinga}, \citenamefont {Scarano}, \citenamefont {Wieneke},\ and\
  \citenamefont {van Oudheusden}}]{elsinga2006tomographic}%
  \BibitemOpen
  \bibfield  {author} {\bibinfo {author} {\bibfnamefont {G.~E.}\ \bibnamefont
  {Elsinga}}, \bibinfo {author} {\bibfnamefont {F.}~\bibnamefont {Scarano}},
  \bibinfo {author} {\bibfnamefont {B.}~\bibnamefont {Wieneke}},\ and\ \bibinfo
  {author} {\bibfnamefont {B.~W.}\ \bibnamefont {van Oudheusden}},\ }\bibfield
  {title} {\bibinfo {title} {Tomographic particle image velocimetry},\
  }\href@noop {} {\bibfield  {journal} {\bibinfo  {journal} {Exp. Fluids}\
  }\textbf {\bibinfo {volume} {41}},\ \bibinfo {pages} {933} (\bibinfo {year}
  {2006})}\BibitemShut {NoStop}%
\bibitem [{\citenamefont {Jiang}(2024)}]{jiang2024priv}%
  \BibitemOpen
  \bibfield  {author} {\bibinfo {author} {\bibfnamefont {L.}~\bibnamefont
  {Jiang}},\ }\href@noop {} {\bibinfo {title} {{Private {C}ommunication on the
  ``Rotation of anisotropic particles in {R}ayleigh--{B}{\'e}nard turbulence''
  published in J. Fluid Mech., Volume 901, Page A8, 2020.}}} (\bibinfo {year}
  {2024})\BibitemShut {NoStop}%
\bibitem [{\citenamefont {Jiang}\ \emph {et~al.}(2020)\citenamefont {Jiang},
  \citenamefont {Calzavarini},\ and\ \citenamefont {Sun}}]{jiang2020rotation}%
  \BibitemOpen
  \bibfield  {author} {\bibinfo {author} {\bibfnamefont {L.}~\bibnamefont
  {Jiang}}, \bibinfo {author} {\bibfnamefont {E.}~\bibnamefont {Calzavarini}},\
  and\ \bibinfo {author} {\bibfnamefont {C.}~\bibnamefont {Sun}},\ }\bibfield
  {title} {\bibinfo {title} {{Rotation of anisotropic particles in
  {R}ayleigh--{B}{\'e}nard turbulence}},\ }\href@noop {} {\bibfield  {journal}
  {\bibinfo  {journal} {J. Fluid Mech.}\ }\textbf {\bibinfo {volume} {901}},\
  \bibinfo {pages} {A8} (\bibinfo {year} {2020})}\BibitemShut {NoStop}%
\bibitem [{\citenamefont {Cleveland}(1979)}]{cleveland1979robust}%
  \BibitemOpen
  \bibfield  {author} {\bibinfo {author} {\bibfnamefont {W.~S.}\ \bibnamefont
  {Cleveland}},\ }\bibfield  {title} {\bibinfo {title} {Robust locally weighted
  regression and smoothing scatterplots},\ }\href@noop {} {\bibfield  {journal}
  {\bibinfo  {journal} {J. Am. Stat. Assoc.}\ }\textbf {\bibinfo {volume}
  {74}},\ \bibinfo {pages} {829} (\bibinfo {year} {1979})}\BibitemShut
  {NoStop}%
\bibitem [{\citenamefont {Lynch}\ and\ \citenamefont
  {Park}(2017)}]{lynch2017modern}%
  \BibitemOpen
  \bibfield  {author} {\bibinfo {author} {\bibfnamefont {K.~M.}\ \bibnamefont
  {Lynch}}\ and\ \bibinfo {author} {\bibfnamefont {F.~C.}\ \bibnamefont
  {Park}},\ }\href@noop {} {\emph {\bibinfo {title} {Modern robotics}}}\
  (\bibinfo  {publisher} {Cambridge University Press},\ \bibinfo {year}
  {2017})\ p.~\bibinfo {pages} {77}\BibitemShut {NoStop}%
\end{thebibliography}%

\end{document}